\documentclass[floatfix,letterpaper,longbibliography,nofootinbib,aps,pre,preprint,superscriptaddress,showkeys,showpacs,12pt]{revtex4-1}
\usepackage{caption}
\usepackage[caption=false]{subfig}
\usepackage{amsfonts}
\usepackage{amsmath, amsthm, amssymb}
\usepackage{mathtools}
\usepackage{graphicx}
\usepackage{float}
\usepackage{breqn}
\usepackage{hyperref}
\usepackage{bm}
\usepackage{url}
\usepackage{placeins}
\newcommand\numberthis{\addtocounter{equation}{1}\tag{\theequation}}

\numberwithin{equation}{section}

\begin{document}
\title{Tensor Green's Function Evaluation in Arbitrarily Anisotropic, Layered Media using Complex-Plane Gauss-Laguerre Quadrature}
\date{\today}
\author{Kamalesh Sainath}
\email{sainath.1@osu.edu}
\author{Fernando L. Teixeira}
\email{teixeira@ece.osu.edu}
\affiliation{The Ohio State University: ElectroScience Laboratory}
\altaffiliation[Address: ]{1330 Kinnear Road, Columbus, Ohio, USA 43212}

\begin{abstract}
\noindent
We discuss the application of Complex-Plane Gauss-Laguerre Quadrature (CGLQ) to efficiently evaluate two-dimensional Fourier integrals arising as the solution to electromagnetic fields radiated by elementary dipole antennas embedded within planar-layered media with arbitrary material parameters. More specifically, we apply CGLQ to the long-standing problem of rapidly and efficiently evaluating the semi-infinite length ``tails" of the Fourier integral path while simultaneously and robustly guaranteeing absolute, exponential convergence of the field solution despite diversity in the doubly anisotropic layer parameters, source type (i.e., electric or equivalent magnetic dipole), source orientation, observed field type (magnetic or electric), (non-zero) frequency, and (non-zero) source-observer separation geometry. The proposed algorithm exhibits robustness despite unique challenges arising for the fast evaluation of such two-dimensional integrals. Herein, we (1) develop the mathematical treatment to rigorously evaluate the tail integrals using CGLQ and (2) discuss and address the specific issues posed to the CGLQ method when anisotropic, layered media are present. To empirically demonstrate the CGLQ algorithm's computational efficiency, versatility, and accuracy, we perform a convergence analysis along with two case studies related to (a) modeling of electromagnetic resistivity tools employed in geophysical prospection of layered, anisotropic Earth media and (b) validating the ability of isoimpedance substrates to enhance the radiation performance of planar antennas placed in close proximity to metallic ground planes.
\end{abstract}

\maketitle
\section{Introduction}
\begin{figure}[H]
\centering
\subfloat[\label{GPR}]{\includegraphics[width=3.5in]{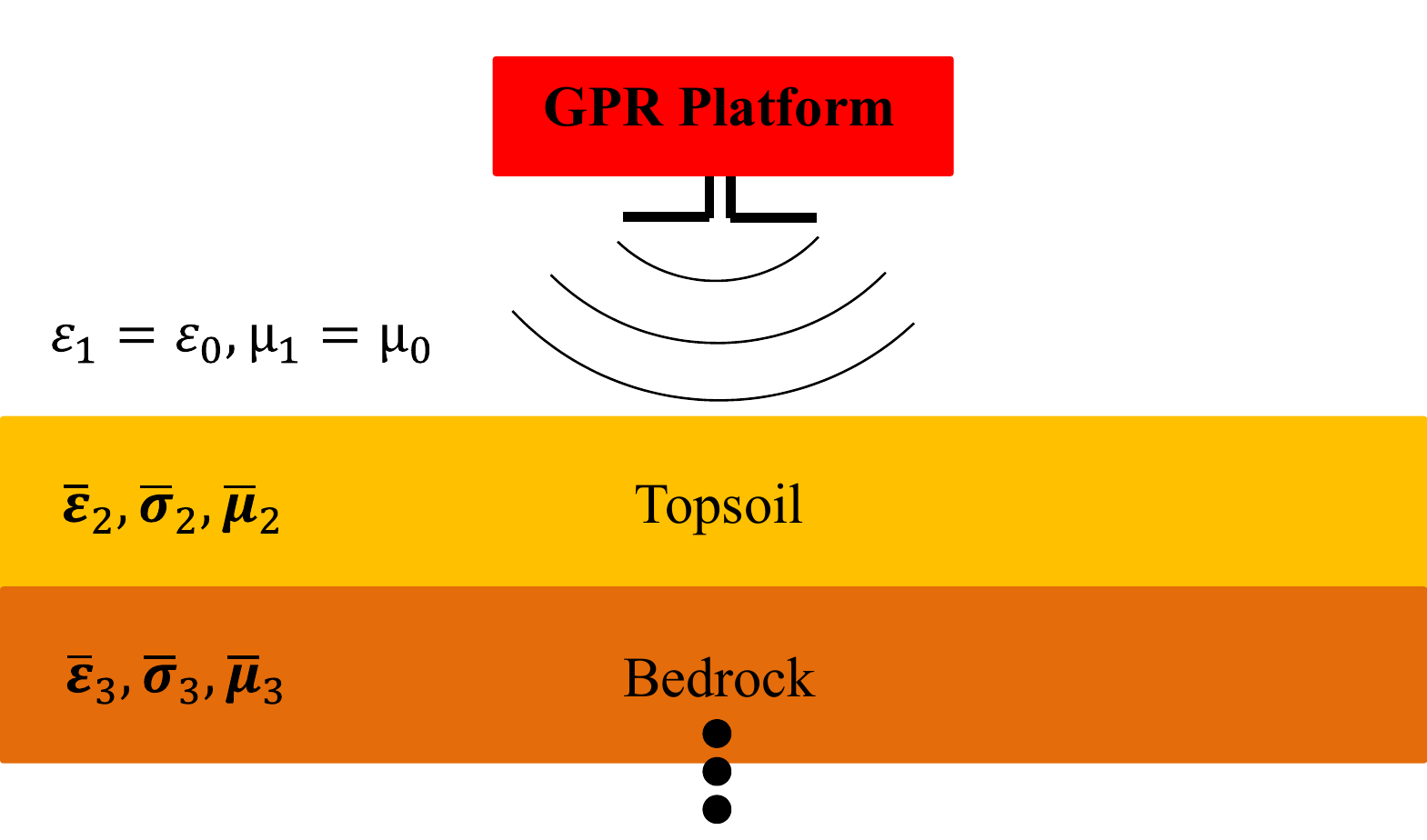}}

\subfloat[\label{ionosphere}]{\includegraphics[width=3.5in]{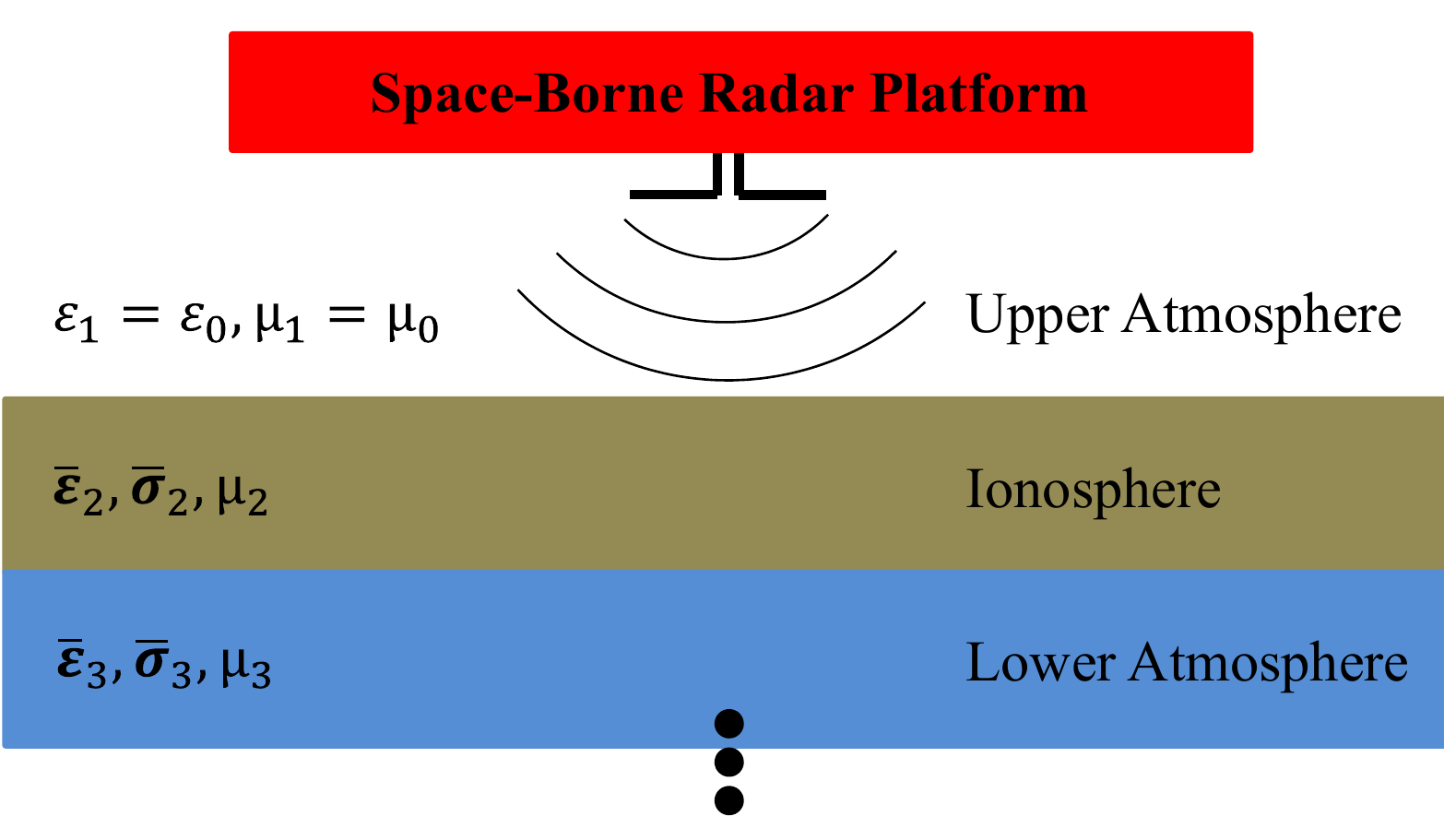}}
\caption{\small Schematic illustration of two application areas frequently encountering environments well-approximated and modeled as planar-layered media containing one or more anisotropic layers. Figure \ref{GPR} illustrates usage of ground-penetrating radar (GPR) in subsurface material profile retrieval (i.e., an example of solving the inverse EM problem), while Figure \ref{ionosphere} illustrates radio-wave propagation through and distortion by an inhomogeneous, dispersive atmosphere potentially containing one or more anisotropic layers. Note: Contrary to what Figure \ref{Apps} suggests, our algorithm also admits arbitrarily anisotropic material parameters in layer one.}
\label{Apps}
    \end{figure}

A long-standing need exists to efficiently, robustly, and accurately solve time-harmonic electromagnetic (EM) radiation and scattering problems in layered media \cite{sommerfeld}. Applications regularly encountering problem scenarios approximated by planar-layered, anisotropic media include hydrocarbon well-logging using radar and induction instruments \cite{sofia,wei,anderson1,anderson2,howard,zhdanov,wang,moran1,HOLee2012,HOLee2007,rabinovich1,sato1}, analysis and design of both microwave circuits and antennas \cite{mosig2,pozar,pozar2}, plasma physics \cite{paulus,felsen}, atmospheric studies \cite{jehle}, ground penetrating radar (GPR) \cite{lambot1,lambot2}, and optical field manipulation \cite{jain}. Illustrations of application areas requiring algorithms with such features can be found in Figure \ref{Apps} below and Figure 1 of \cite{sainath2}. Computational cost is a critical aspect in many cases, such as when attempting to solve inverse EM problems, due to the need for solving the forward EM problem many times to effect successful extraction of the desired environmental parameters \cite{spagnolini1}. To address the simultaneous needs to solve layered-media problems both rigorously \emph{and} efficiently, pseudo-analytic approaches, which consist of posing the EM field solution as an inverse Fourier-type integral that synthesizes the required Green's Tensor components as a spectral superposition of modal fields (e.g., characteristic plane waves), often represent the preferred numerical solution method~\cite{chew,felsen}. Such integrals typically assume the form, for some tensor Green's function component $\Psi(\bold{r})$, as either a two-dimensional Fourier integral \eqref{psi1} or one-dimensional Fourier-Hankel integral \eqref{psi2}\footnote{The formulation presented herein is readily applicable to Sommerfeld integrals (i.e., Fourier-Bessel transforms \cite{sommerfeld}) and also to fields evaluated in cylindrically-layered media employing similar integral representations \cite{chew}[Ch. 2,4].}:
\begin{align}
\Psi_1(\bold{r})&\sim\iint_{C_1}\tilde{\Psi}_1(k_x,k_y)\mathrm{e}^{ik_x(x-x')+ik_y(y-y')+i\tilde{k}_z(z-z')}\mathrm{d}k_x \mathrm{d}k_y \numberthis \label{psi1} \\
\Psi_2(\bold{r})&\sim\int_{C_2}\tilde{\Psi}_2(k_{\rho})H_n^{(1)}(k_{\rho}|\rho-\rho'|)\mathrm{e}^{i\tilde{k}_z(z-z')}\mathrm{d}k_{\rho}\numberthis \label{psi2}
\end{align}
where $H_n^{(1)}(k_{\rho}|\rho-\rho'|)$ is the $n$th order Hankel function of the first kind, representing an outgoing cylindrical wave\footnote{The exp($-i\omega t$) time harmonic convention is assumed and suppressed throughout.}, $\tilde{\Psi}$ is the spectral domain analog of the space domain function $\Psi$, $\bold{k}=(k_x,k_y,\tilde{k}(k_x,k_y))$ is the wave vector, $k_{\rho}=\sqrt{k_x^2+k_y^2}$, $\rho=\sqrt{x^2+y^2}$, $\bold{r}=(x,y,z)$ is the field observation point, and $\bold{r}'=(x',y',z')$ is the source location. Note that the first integral \eqref{psi1}, but not the second one \eqref{psi2}, is suitable for evaluating fields in either isotropic or arbitrarily anisotropic\footnote{We assume, to ensure the completeness of the plane wave basis, that the material tensors are diagonalizable. However, this constraint is not limiting in practical problems since all natural media are characterized by diagonalizable permittivity and permeability tensors.} planar-layered media and thus represents the class of integrals we examine further\footnote{The methodology developed below can be applied to integrals of the form \eqref{psi2}, and by extension to Sommerfeld integrals via an appropriate transformation \cite{chew}[Ch. 2], by setting the optimal path detour angle as $\gamma=\mathrm{tan}^{-1}(|\rho-\rho'|/|z-z'|)$ in the case of planar-layered media (see Section \ref{evan}). A similar formula for $\gamma$ applies for cylindrically-layered media.}.

Despite the rigor of the solution method provided by such integrals, in practice when subject to direct numerical evaluation using a path on or near the real axis, the integrand may exhibit highly oscillatory or weakly convergent behavior for various types of source-observer separation geometries $\bold{r}-\bold{r}'$ of interest (i.e., $|\rho-\rho'|=\sqrt{(x-x')^2+(y-y')^2} \gg 1$ or $0 \leq |z-z'|\ll 1$ resp.).
To address these challenges, various approaches have been developed over the years. On one side are techniques aimed at circumventing the need to perform direct numerical integration altogether. Prominent among this class of methods are closed-form asymptotic solutions \cite{chew}[Ch. 2]\cite{felsen} and image methods \cite{mich1,mich5,caboussat}. Asymptotic techniques typically feature geometry-specific applicability and accuracy depending on an asymptotic value of one or more parameters (e.g., frequency, observation distance, etc.) \cite{chew}[Ch. 2]. On the other hand, image methods are known to typically lack robust error-control mechanisms \cite{mosig7} in addition to also exhibiting geometry-specific applicability \cite{mich1}. In contrast to these techniques, a different strategy consists in attempting the (efficient) direct numerical integration. Among these techniques are the so-called ``weighted average"-type techniques \cite{mich2,mich1,mosig1,mosig3,sainath}, which belong to the broader family of scalar Levin sequence transforms \cite{homeier}. These methods treat the full, non-truncated Sommerfeld, Fourier-Hankel and Fourier tail integrals as a sum of integrals, each of whose paths span a finite section of the tail, and devise a ``weighted average" formulation that, in effect, aptly guesses, compensates for, and thereby reduces the truncation error associated with evaluating only a finite section of the integration path tail. A recent extension to this method developed in \cite{sainath2}, and denoted as the ``Complex-Plane Method of Weighted Averages" (CPMWA), consists of (1) deforming the Fourier integral tail path into a linear path impinging into the upper-half of the complex plane, (2) partitioning the deformed path into finite-length intervals, and (3) adaptively taking weighted averages of a successively greater number of estimations of the non-truncated tail integral.
Through validation and convergence studies, it was demonstrated \cite{sainath2} that this strategy could rigorously guarantee absolute, exponential-cum-algebraic convergence for a wide range of planar-layered problems, a significant improvement over the (real-axis) extrapolation methods employed in the past~\cite{mich2,sainath,mosig1}.

We should also note the possibility of numerically evaluating the integral along the Steepest Descent Path (SDP). However, the possibility of intersecting and (or) deforming past critical points on the complex-plane\footnote{That is, branch points, branch cuts, or poles.}, along with (1) the requirement to identify and integrate through the saddle point, (2) the book-keeping necessary to track all critical-point-crossing occurrences, and (3) having to analytically account for these problem-dependent critical-point-crossings at the post-integration stage makes such a method less desirable. Due to similar book-keeping needs and problem-dependent characteristics, we also avoid use of the integration path suggested in \cite{mosig5}.

Despite its robustness, the CPMWA still presents some drawbacks associated with (1) the large number of integrand evaluations necessary to evaluate the full integral tail, (2) the need to pre-compute the set of weights required for an adaptive implementation~\cite{sainath2}, as well as (3) residual aliasing and numerical stability considerations. In particular, the need to mitigate aliasing caused by unduly long extrapolation region intervals can force the choice of suboptimal path deformation detour angles (see \cite{sainath2} for details). Additionally, the efficiency and numerical stability of all the extrapolation methods discussed above implicitly relies upon the oscillatory behavior of the integrand~\cite{mich1,mosig1,sainath,sainath2}. Indeed, this oscillatory characteristic of the integrand was assumed in \cite{sainath2} due to the (practical) inability to (in general) construct a rigorous, mode-independent Constant-Phase Path (CPP) as a result of the presence of different locations for the critical points according to the individual anisotropic layer parameters, layer thicknesses, and so on. However, when the numerical integration does occur along or very near to the asymptotic CPP, computation of the $k_x$ integral weights becomes a numerically unstable procedure for $0 \leq |x-x'| \ll 1$ \cite{sidi}, which necessitates an ad-hoc adjustment to the MWA-type weight computation methodology\footnote{An analogous statement holds for the weights used to compute the $k_y$ tail integrals.}. Therefore, a new method eliminating the (1) excessive integrand evaluations and pre-computation of multiple weight sets, (2) potential numerical instability and subsequent need for ad-hoc adjustment of the weight computation method, and (3) artificial (i.e., algorithm-dependent) added constraints placed upon the path deformation detour angles to mitigate aliasing and the number of pre-computed weight sets, and instead offering a direct integration procedure with minimal integrand evaluations, no required pre-computation and use of weight sets potentially resulting from an ad-hoc computation scheme, and minimal constraints imposed upon the departure angles\footnote{The term ``minimal constraints" refers to those constraints imposed by the fundamental behavior of the wave dynamics solution as manifest in a Fourier, Fourier-Hankel, or Sommerfeld integral representation.}, while simultaneously guaranteeing absolute, exponential convergence for all ranges of anisotropic, planar-layered problems is highly desirable.

The solution method introduced here to effect these changes is the complex-plane extension of Gauss-Laguerre Quadrature (CGLQ) \cite{davies,sagar}, which in its traditional form (i.e., integration along the real axis) approximates semi-infinite range integrals of the form
 \begin{equation} \int\limits_0^{\infty} \mathrm{e}^{-x}f(x)\mathrm{d}x \end{equation}
 via an order-$P$ numerical quadrature formula $\sum_{m=1}^{P}f(x_m)w_m$, where both the nodes $\{x_m\}$ and weights $\{w_m\}$ are real valued. On the other hand for a general path deformation into the complex plane, parameterized in terms of spanning the semi-infinite range of a real-valued variable, the nodes and weights can both be complex-valued. The deformed path we decide to use is identical in shape to that shown in Figure 2 of \cite{sainath2}, and (ideally) spans (asymptotically) the CPP\footnote{As pointed out in \cite{davies}, the CPP is \emph{not} necessarily equivalent to the SDP \cite[Ch.\ 2]{chew}. In particular, we note that (1) the presence of a saddle point, through which the SDP would proceed, is neither stipulated nor solved for here, and (2) no asymptotic dependence in regards to the observation point is assumed or implied in our present formulation.} along which the exponential phase factors exhibit no oscillation while simultaneously imparting maximum exponential decay to the integrand \cite{davies}. As a result, the exponential decay combined with minimized integrand oscillation makes this integral type an ideal candidate for accurate \emph{and} efficient numerical evaluation by CGLQ.

We note, however, that in the present CGLQ method one removes non-adaptive integration path sub-division, which was used in prior MWA variants \cite{mich1,mich2,mosig1,sainath,sainath2} to increase tail integral accuracy via limiting integrand oscillation. Instead, one now relies solely upon the sufficiently well-behaved nature of the integrand $f(x)$ along the deformed path to facilitate its interpolation via Laguerre polynomials, along with adaptively refining the solution using successively higher-order CGLQ quadrature rules (i.e., $p$ refinement). To minimize any fast integrand variations and thereby facilitate successfully modeling $f(x)$ via these Laguerre polynomials, undesirable integrand oscillations that (dominantly) arise from the exponential complex-phase factors of the form exp($ik_x\Delta x+ik_y\Delta y+i\tilde{k}_{z}\Delta z$) are suppressed here. Note that since we initially perform adaptive $hp$ integration refinement within and sufficiently past the neighborhood of any critical points near the real axis, we assume the critical points themselves do not cause appreciably abrupt variations of $f(x)$ along the tail integral path \cite{mich2,mosig1}. The validation results and convergence study presented here indicate that major gains in both computational efficiency and accuracy robustness, with respect to diverse problem parameters, are realized with only a marginal penalty in accuracy compared to CPMWA.

Before proceeding, we remark that (akin to \cite{sainath2}) it is assumed that one has already performed an azimuthal basis rotation such that in the rotated basis $x-x'=\Delta x = y-y' = \Delta y \geq 0$ while $-\infty < \left(z-z'=\Delta z\right)< \infty $, where $\bold{r}=(x,y,z)$ is the observation point and $\bold{r}'=(x',y',z')$ is the dipole source location in the rotated basis. This rotation is performed to ensure absolute, exponential convergence of both the outer \emph{and} inner integral regardless of the transverse source-observer separation geometry $\rho-\rho'$ while streamlining the formulation dictating the shape of the integration path. Knowledge of all required vector and tensor field transformations done as part of the azimuthal basis rotation is implicitly assumed and not discussed further herein.
\section{Formulation}
\subsection{Propagation Spectra Contribution}
Henceforth we discuss the two-dimensional Fourier integral, rather than the one-dimensional Fourier-Hankel integral, to raise and address specific concerns regarding the former. To this end, let $k_x$ and $k_y$ be the inner and outer integration variables (resp.). Any discussion pertaining to the inner integral, which we assume is being evaluated for a fixed $k_y$ value $k_{y0}$, applies analogously to the outer integral (and vice-versa) unless explicitly stated otherwise. First we briefly summarize treatment of the propagation spectra contributions to the observed field, which mirrors that in \cite{sainath2} (see \cite{sainath} for details on evaluating the integrand), before proceeding to the primary content of this article.

To robustly estimate and avoid the region wherein critical points may lie near the real axis, we employ (1) a conservative estimate
for the multilayered environment's ``effective refractive index" $n^+$ \cite{sainath} and (2) a trapezoidal detour, terminating at $k_x=\pm P_k$ on the real axis, that is parameterized identically to its counterpart in \cite{sainath2} except for setting $P_k=k_0\left(n^+ + 2\right)$, where $k_0=\omega/c$ is the free space wave number, $\omega=2\pi f$ is the angular frequency of radiation, and $c$ is the speed of light in free space\footnote{The presence of the constant $T_0$ in the CGLQ method, as seen upon examining the similar propagation spectrum evaluation methodology in \cite{sainath2}, is simply to bound the detour height.}. Adjoined to this trapezoidal path are real-axis segments spanning the interval $\left(-\xi_1\leq \mathrm{Re}[k_x] \leq -P_k\right) \cup \left(P_k\leq \mathrm{Re}[k_x] \leq \xi_1\right) $; see \cite{sainath2} for calculating $\xi_1$. Within the region $\left(-\xi_1 \leq \mathrm{Re}[k_x] \leq \xi_1 \right)$, labeled herein as "Propagation Region," an error-controllable numerical integration is done through a multi-level $hp$ adaptive refinement. On the other hand, the region containing the integral tails is denoted as ``Evanescent Region"\footnote{Note that these two labels are loosely employed; indeed, there exists (in general) no sharp boundary in wave number space delineating propagating modes from evanescent modes in the presence of planar inhomogeneity and lossy layers.}.
\subsection{\label{evan}Evanescent Spectra Contribution}
The geometry of the path and spectral domain partition used here follows \cite{sainath2}; see Figure 2a and 3 therein for illustrations. From \cite{sainath2}, it was determined that the optimal departure angle $\gamma$ to asymptotically maximize decay of the complex exponential exp($ik_x\Delta x+i\tilde{k}^{\pm}_z\Delta z$) is given by $\gamma=\mathrm{tan}^{-1}\left(\Delta x/ \Delta z\right) $\footnote{As in \cite{sainath2}, we estimate the departure angle assuming the environment were homogeneous and isotropic. Under this approximation one has $\tilde{k}^{\pm}_z=\pm \sqrt{k^2-k_x^2-k_y^2}$, where $k$ is the characteristic wave number of the medium, while for large $|k_x|$ one has the asymptotically-valid relations, i.e., for $|k_x|\gg 0$, $\tilde{k}^{\pm}_z \to \pm ik_x$ (Re[$k_x]\geq 0$) and $\tilde{k}^{\pm}_z \to \mp ik_x$ (Re[$k_x]\leq 0$).}; similarly, choosing $\alpha=\mathrm{tan}^{-1}\left(\Delta y/ \Delta z\right)$ asymptotically maximizes decay of the complex exponential exp($ik_y\Delta y+i\tilde{k}^+_z\Delta z$).

Upon setting $t_o^+=\cos{\gamma^+}+i\sin{\gamma^+}$, $t_o^-=\cos{\gamma^-}-i\sin{\gamma^-}$, and parameterizing the tail integral path as
\begin{equation}
k_x=\begin{cases} \xi_1+t_o^+r_x ,& r_x>0 \\ -\xi_1+t_o^-r_x ,& r_x<0 \end{cases}
\end{equation}
  the half-tail integrals $\Psi_e^{\pm}$ in the $\pm \mathrm{Re}[k_x]$ half-planes, corresponding physically to evanescent spectra contributions to the observed field, asymptotically behave as (resp.) \cite{sainath2}
\begin{align}
\Psi_e^+&=t_o^+\mathrm{e}^{ik_{y0}\Delta y+\xi_1(i\Delta x-\Delta z)}\int\limits_{0}^{\infty} \tilde{g}(\xi_1+r_xt_o^+,k_{y0})\mathrm{e}^{r_x t_o^+(i\Delta x -\Delta z)}\mathrm{d}r_x \numberthis \label{tailint1a} \\
\Psi_e^-&=t_o^-\mathrm{e}^{ik_{y0}\Delta y-\xi_1(i\Delta x+\Delta z)}\int\limits_{-\infty}^{0} \tilde{g}(-\xi_1+r_xt_o^-,k_{y0})\mathrm{e}^{r_x t_o^-(i\Delta x +\Delta z)}\mathrm{d}r_x \numberthis \label{tailint1b}
\end{align}
where without loss of generality it is assumed that $\Delta z=z-z' \geq 0$. Recalling the definitions of $t_o^{\pm}$, setting $\tau^{\pm}=\Delta x\sin \gamma^{\pm}+\Delta z\cos \gamma^{\pm}$ \footnote{We use the convention $\tau^{\pm}=\Delta x\sin \gamma^{\pm}+\Delta z\cos \gamma^{\pm}$ to compactly denote the relations $\tau^+=\Delta x\sin \gamma^++\Delta z\cos \gamma^+$ and $\tau^-=\Delta x\sin \gamma^-+\Delta z\cos \gamma^-$ simultaneously. An analogous comment applies for other expressions bearing this plus-minus superscript type of convention.}, and defining $\beta^{\pm}=\Delta x\cos \gamma^{\pm}-\Delta z\sin \gamma^{\pm}$, we find that \eqref{tailint1a}-\eqref{tailint1b} asymptotically become
\begin{align}
\Psi_e^+&=t_o^+\mathrm{e}^{ik_{y0}\Delta y+\xi_1(i\Delta x-\Delta z)}\int\limits_{0}^{\infty} \tilde{g}(\xi_1+r_xt_o^+,k_{y0})\mathrm{e}^{-r_x\left(\tau^+ - i\beta^+\right)}\mathrm{d}r_x \numberthis \label{tailint2a} \\
\Psi_e^-&=t_o^-\mathrm{e}^{ik_{y0}\Delta y-\xi_1(i\Delta x+\Delta z)}\int\limits_{-\infty}^{0} \tilde{g}(-\xi_1+r_xt_o^-,k_{y0})\mathrm{e}^{r_x\left(\tau^- + i\beta^-\right)}\mathrm{d}r_x\numberthis \label{tailint2b}
\end{align}
By making the change of variable $r_x=-r_x'$ in \eqref{tailint2b} and subsequently dropping the prime, one has
\begin{align}
\Psi_e^+&=t_o^+\mathrm{e}^{ik_{y0}\Delta y+\xi_1(i\Delta x-\Delta z)}\int\limits_{0}^{\infty} \tilde{g}(\xi_1+r_xt_o^+,k_{y0})\mathrm{e}^{-r_x\left(\tau^+ - i\beta^+\right)}\mathrm{d}r_x \numberthis \label{tailint3a} \\
\Psi_e^-&=t_o^-\mathrm{e}^{ik_{y0}\Delta y-\xi_1(i\Delta x+\Delta z)}\int\limits_{0}^{\infty} \tilde{g}(-\xi_1-r_xt_o^-,k_{y0})\mathrm{e}^{-r_x\left(\tau^- + i\beta^-\right)}\mathrm{d}r_x\numberthis \label{tailint3b}
\end{align}
Next, by making the substitution $r_x^{\pm}=r_x\tau^{\pm}$, subsequently dropping the ``$\pm$" superscripts in $r_x^{\pm}$, and defining $l^{\pm}=t_o^{\pm}/\tau^{\pm}$, one obtains the following pair of integrals suitable for evaluation by complex-plane Gauss-Laguerre quadrature:
\begin{align}
\Psi_e^+&=l^+\mathrm{e}^{ik_{y0}\Delta y+\xi_1(i\Delta x-\Delta z)}\int\limits_{0}^{\infty} \mathrm{e}^{-r_x}\tilde{g}(\xi_1+l^+r_x,k_{y0})\mathrm{e}^{ir_x\beta^+/{\tau}^+}\mathrm{d}r_x \numberthis \label{tailint4a} \\
\Psi_e^-&=l^-\mathrm{e}^{ik_{y0}\Delta y-\xi_1(i\Delta x+\Delta z)}\int\limits_{0}^{\infty} \mathrm{e}^{-r_x}\tilde{g}(-\xi_1-l^-r_x,k_{y0})\mathrm{e}^{-ir_x\beta^-/{\tau}^-}\mathrm{d}r_x\numberthis \label{tailint4b}
\end{align}
where the $k_x$ plane nodes and weights ($\{k_{xp}\}$ and $\{w_{xp}\}$) are related to the real-valued $r_x$ plane nodes and weights ($\{r_{xp}\}$ and $\{w_{rp}\}$), used to evaluate \eqref{tailint4a} and \eqref{tailint4b}, as (resp.)
\begin{align} k_{xp}&=\xi_1+l^+r_{xp}, \ w_{xp}=w_{rp} \\
 k_{xp}&=-\xi_1-l^-r_{xp}, \ w_{xp}=w_{rp}
 \end{align}
 such that one can now efficiently compute \eqref{tailint4a}-\eqref{tailint4b}, with \emph{zero} tail integral truncation error, as
 \begin{align}
\Psi_e^+&\sim l^+\mathrm{e}^{ik_{y0}\Delta y+\xi_1(i\Delta x-\Delta z)}\sum_{p=1}^{P}\mathrm{e}^{ir_x\beta^+/{\tau}^+}\tilde{g}(\xi_1+l^+r_{xp},k_{y0})w_{rp} \\
\Psi_e^-&\sim l^-\mathrm{e}^{ik_{y0}\Delta y-\xi_1(i\Delta x+\Delta z)}\sum_{p=1}^{P}\mathrm{e}^{-ir_x\beta^-/{\tau}^-}\tilde{g}(-\xi_1-l^-r_{xp},k_{y0})w_{rp}
 \end{align}
using a $P$-point Gauss-Laguerre numerical quadrature formula.

However, since \eqref{tailint4a}-\eqref{tailint4b} is only \emph{asymptotically} true, there will be a residual error associated with approximating $i\tilde{k}^+_z$ as $-\cos{\gamma^{\pm}}r_x-\xi_1 \mp i\sin{\gamma^{\pm}}r_x$, where the top and bottom signs of this expression's ``$\pm$" and ``$\mp$" symbols hold for $r_x>0$ and $r_x<0$ (resp.). Therefore, in having extracted the term $-r_x\Delta z\cos{\gamma^{\pm}}$ in \eqref{tailint3a}-\eqref{tailint3b} to create the exponential Laguerre polynomial weight factor $\mathrm{exp}\left(-r_x\tau^{\pm}\right)$, to ensure analytical exactness in the formulation one must account for this extraction via ``adding back in" the term $+r_x\Delta z\cos{\gamma^{\pm}}$ that is expected to (asymptotically) cancel with Re[$i\tilde{k}^{\pm}_z\Delta z$] up to the factor $\left(-\xi_1 \mp i\sin{\gamma^{\pm}}r_x\right)\Delta z$ \footnote{When planar inhomogeneity or anisotropy is involved, naturally the extent of asymptotic cancelation that occurs in reality can exhibit great variation with respect to mode type and the media involved. To account for such uncertainty, one can robustly mitigate exponentially rising terms via placing the natural logarithm of the CGLQ numerical quadrature weights in the argument of the exponential, seen in the integrand of \eqref{tailint5a}, prior to evaluating the exponent.}. Recalling the final variable transform made in deriving \eqref{tailint4a}-\eqref{tailint4b} from \eqref{tailint3a}-\eqref{tailint3b}, one finally arrives at the exact expressions
\begin{equation}\label{tailint5a} \Psi_e^{\pm}=l^{\pm}\mathrm{e}^{ik_{y0}\Delta y \pm i\xi_1\Delta x} \int\limits_{0}^{\infty}\mathrm{e}^{-r_x} \tilde{g}(\pm \xi_1 \pm l^{\pm}r_x,k_{y0})\mathrm{e}^{\Delta z(i\tilde{k}_z^{+}+(r_x/\tau^{\pm})\cos{\gamma^{\pm}})\pm i(r_x/\tau^{\pm})\Delta x\cos{\gamma^{\pm}}}\mathrm{d}r_x
\end{equation}
\subsection{Comments on the Constant Phase Path}
The above analysis shows that the (ideal) detour angle maximizing the integrand's exponential decay is given by $\gamma=\mathrm{\tan}^{-1}\left(\Delta x/ \Delta z\right)$, with the associated function providing the decay asymptotically expressed as exp(-$r_x\sqrt{(\Delta x)^2+(\Delta z)^2}$) in the event of the actual and ideal detour angles coinciding. Furthermore, one can easily show that along this path the phase associated with the complex exponential is (asymptotically) non-varying with respect to $r_x$ \cite{davies}, hence the name ``Constant-Phase Path''. However, in practice one may not actually be able to deform (asymptotically) onto the exact CPP due to the presence of critical points, as well as the necessity to preclude their migration (a) into the second and fourth quadrants of the $k_x$ plane and (b) towards Re$[k_x]=\pm \infty$. Although the impact of these requirements on the detour angles can be mitigated, in both the CPMWA and present CGLQ algorithms, through a suitable partitioning of the integration domain (see Figure 3 in \cite{sainath2}), these requirements still can prevent deforming onto the optimal path that asymptotically maximizes numerical accuracy and convergence speed (compared to other tail path deformation angles). Furthermore, in general two main factors prevent one from even defining a unique, common longitudinal propagation distance traversed by the modal fields (and hence a unique, common CPP associated with the four modal contributions to the observed field \cite{sainath,sainath2}) \cite{chew}[Ch. 2]:
\begin{enumerate}
\item Planar stratification leads to the presence of characteristic modes exhibiting layer-dependent longitudinal propagation constants. Furthermore, due to reflections at layer interfaces, both up-going and down-going modes are typically present, which in general travel different effective longitudinal distances before reaching the observation point $\bold{r}$. The planar stratification also produces, in general, multi-bounce of fields within each slab layer.
\item Anisotropy leads to mode-dependent longitudinal propagation constants along with cross-coupling of characteristic modes at the planar interfaces.
\end{enumerate}
The inability, in practice, to robustly define a unique CPP departure angle and to deform onto the estimated (asymptotically) optimal path leads to unwanted, residual integrand oscillation along the actual integration path. In principle, these residual oscillations are expected to restrict the practical range of applicability within which accurate results can be delivered. Indeed, due to evaluating the semi-infinite tail integrals without path pre-partitioning, the CGLQ algorithm's only error control mechanism for evaluating the evanescent spectra contributions consists of adaptive $p$ refinement. However, the empirical results presented in the following section show, nevertheless, that the CGLQ method delivers excellent accuracy in strong accord with the data both from our previously developed CPMWA algorithm \cite{sainath2} and the results published in \cite{sofia} (se Figures \ref{SofiaFig4Hxx}-\ref{SofiaFig4Hzz}). We emphasize that this excellent agreement manifests despite the presence of multiple anisotropic layers (both uniaxial and biaxial)
in these examples, which in Figures \ref{SofiaFig4Hxx}-\ref{SofiaFig4Hzz} also exhibit significant conductive loss. The algorithm's accuracy is also distinctly manifest in its ability to confirm standard, expected results from the employment of Transformation Optics media (see Figures \ref{CasePic}-\ref{Iso2} below).
\section{\label{results}Validation Results}
To verify the accuracy and efficiency of the proposed CGLQ algorithm, and to numerically assess the impact of the issues considered above, we now exhibit results concerning computation of fields radiated by elementary dipole sources (tensor Greens' function components) embedded in planar-layered, anisotropic media. The results illustrate the algorithm's performance in a wide range of environments with respect to layer material parameters, source-observer geometrical configurations, and a wide range of frequencies spanning 1kHz to 13.56MHz (i.e., \emph{five} decades of frequency).
\subsection{Resistivity Well-Logging: Induction Sondes' Response}
First, we show a data set related to the use of induction-regime electromagnetic sondes for resistivity well-logging (hydrocarbon prospection) in layered geologic formations exhibiting uniaxial or biaxial resistivity in their effective \cite{anderson1,anderson2,rabinovich1} resistivity tensors. To facilitate comparison of accuracy of the computed field solution between the CGLQ and CPMWA methods, we choose the same  set of results used for the CPMWA algorithm in \cite{sainath2}. Details of the simulation problem parameters can be found in \cite{sainath2,sofia} and are summarized here as follows: the induction tool axis dip and strike angles are $\alpha=89^{\circ}$ and $\beta=0^{\circ}$ (resp.), the tool's frequency of electromagnetic emissions is $f$=2MHz, the distance $L$ between the receiver and transmitter loop antennas along the axis of the sonde is 40"=1.016m, the interface partitioning the two-layer formation is located at $D$=0m, the diagonal matrices $R_1$=diag$[100,R_{y'y',1},500]\Omega$m and $R_2$=diag$[1,R_{y'y',2},5]\Omega$m describe the resistivity tensors for layer one (top layer) and layer two (bottom layer) respectively, and we alert the reader to the reversal (versus that in \cite{sofia}) in the resistivity tensor labels assigned, on each of the four pages containing the induction logging plots, between the two plots on the first row and two plots on the third row. For the top row of each plot set in Figures \ref{SofiaFig4Hxx}-\ref{SofiaFig4Hzz} $R_{y'y',1}=200\Omega$m and $R_{y'y',2}=2\Omega$m; similarly, for the middle row in each plot set $R_{y'y',1}=100\Omega$m and $R_{y'y',2}=1\Omega$m while for the bottom row in each plot set $R_{y'y',1}=50\Omega$m and $R_{y'y',2}=0.5\Omega$m. A schematic illustration of the geophysical sonde is depicted in Figure \ref{sonde} below.

Besides Figure \ref{Sofia4e3I} (where there is still acceptable accord for the intended application), we observe excellent agreement between the CGLQ algorithm (blue hatched line curves) and the results  in \cite{sofia} (solid red line curves). Moreover, across \emph{all} the plots in Figures \ref{SofiaFig4Hxx}-\ref{SofiaFig4Hzz} there is a strong accord between the CGLQ and CPMWA (dotted green line) algorithms, suggesting that observed discrepancies versus the data in \cite{sofia} can perhaps trace down to the inaccuracies of the algorithm utilized in generating the initially published reference data \cite{sofia}.

\begin{figure}[H]
\centering
\includegraphics[width=3in]{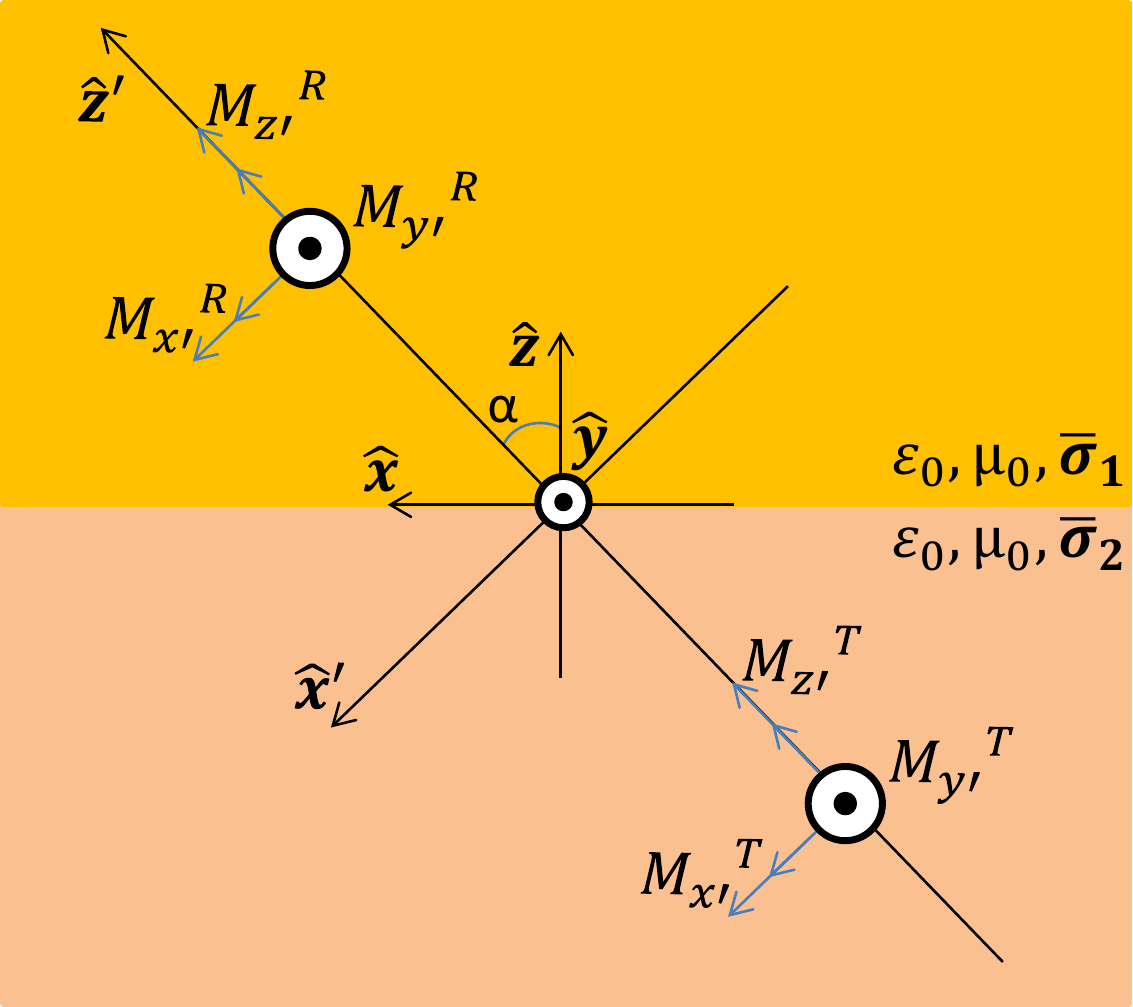}
\caption{\small Schematic description of a standard triaxial electromagnetic sonde, which consists of a system of electrically small loop antennas that are modeled as Hertzian dipoles supporting an equivalent magnetic current (i.e., three orthogonally-oriented, co-located transmitters $M_{x'}^T$, $M_{y'}^T$, and $M_{z'}^T$ spaced a distance of $L$=1.016m from three orthogonally-oriented, co-located receivers $M_{x'}^R$, $M_{y'}^R$, and $M_{z'}^R$) \cite{sofia}. The ``tool coordinate" $x'y'z'$ system, rotated by an angle $\alpha$ with respect to the standard $xyz$ coordinate system, is such that the $z'$ axis is parallel to the ``tool axis" \cite{sofia,moran1}.}
\label{sonde}
\end{figure}
\begin{figure}[H]
\centering
\subfloat[\label{Sofia4a1R}]{\includegraphics[width=2.7in]{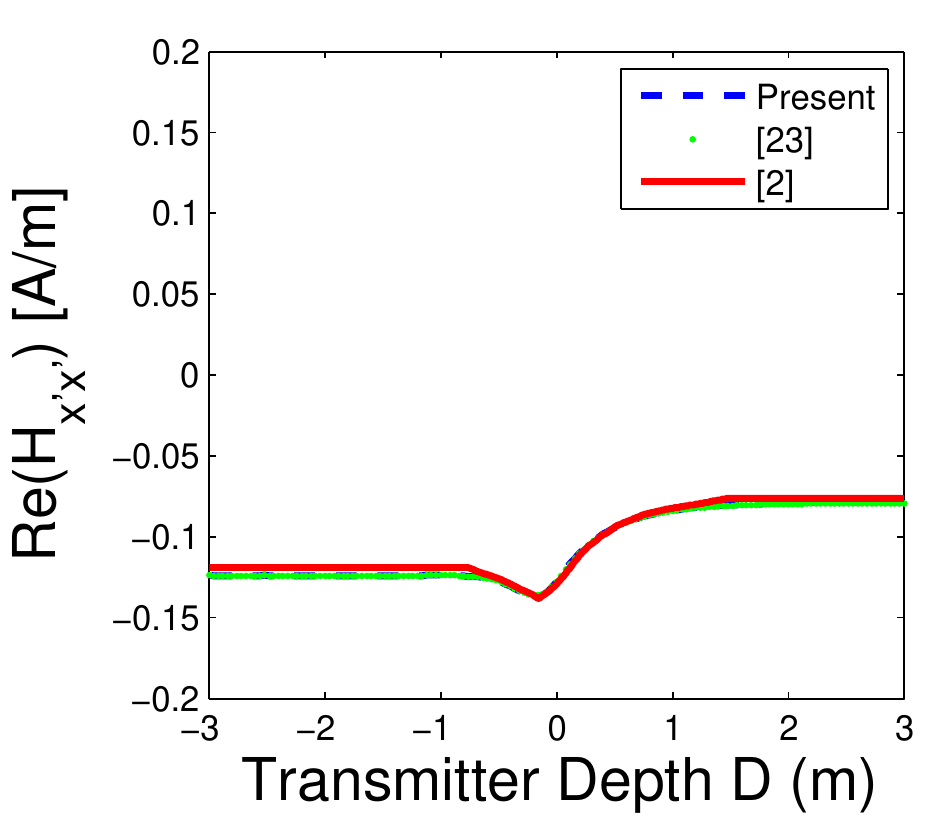}}
\subfloat[\label{Sofia4a1I}]{\includegraphics[width=2.7in]{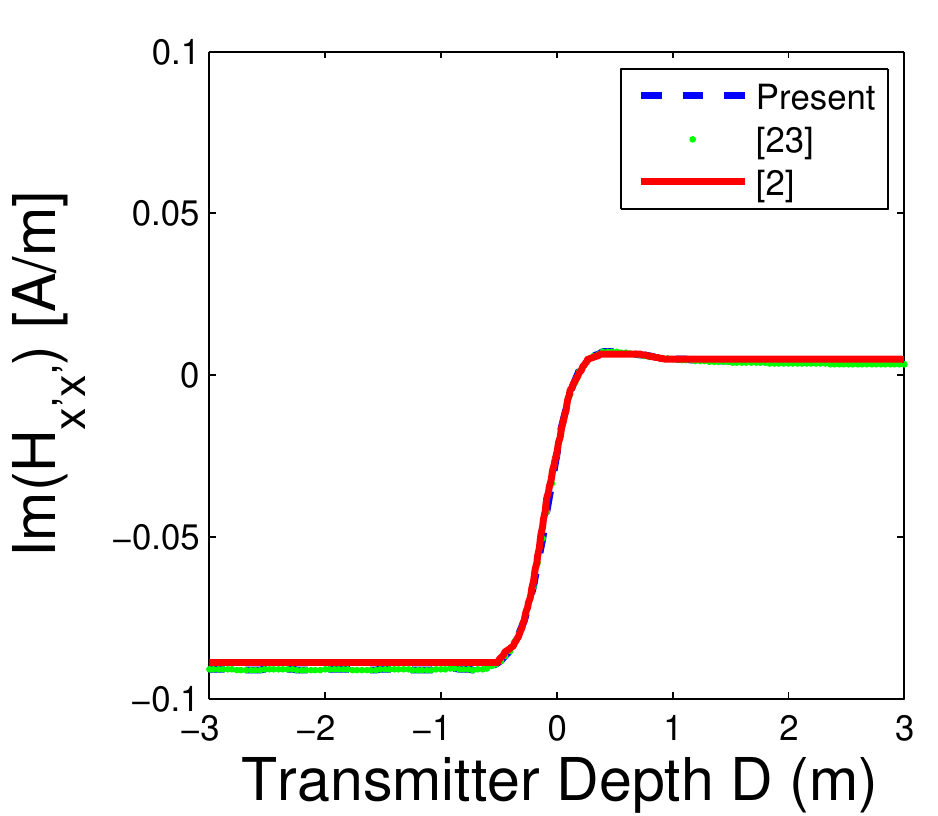}}

\subfloat[\label{Sofia4a2R}]{\includegraphics[width=2.7in]{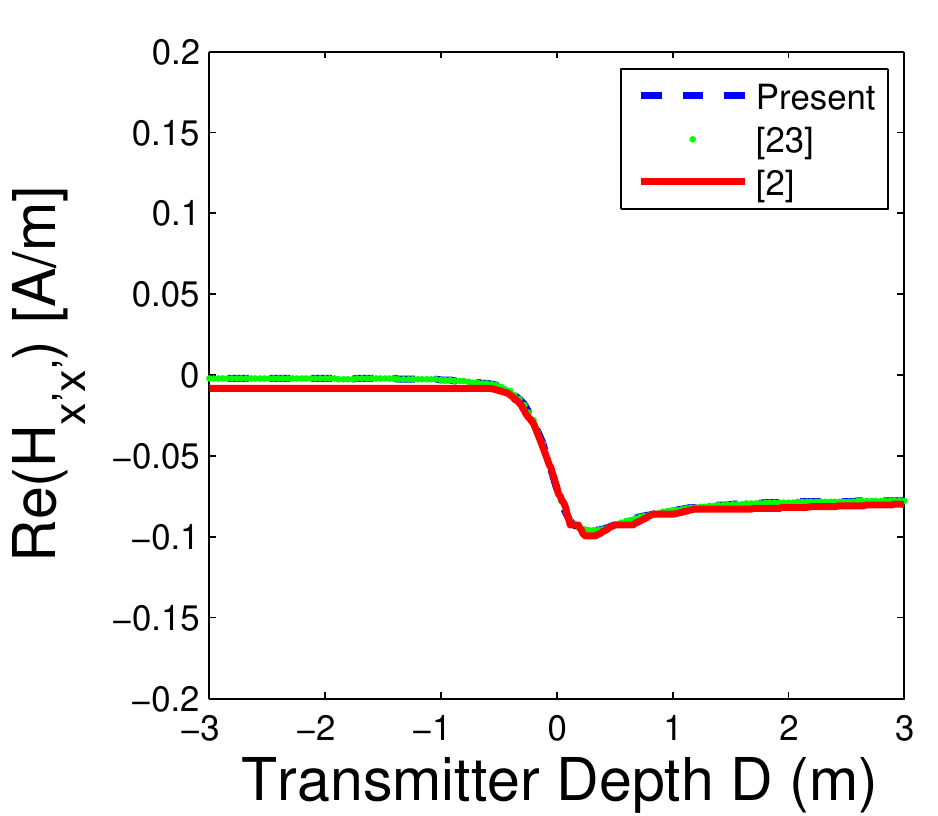}}
\subfloat[\label{Sofia4a2I}]{\includegraphics[width=2.7in]{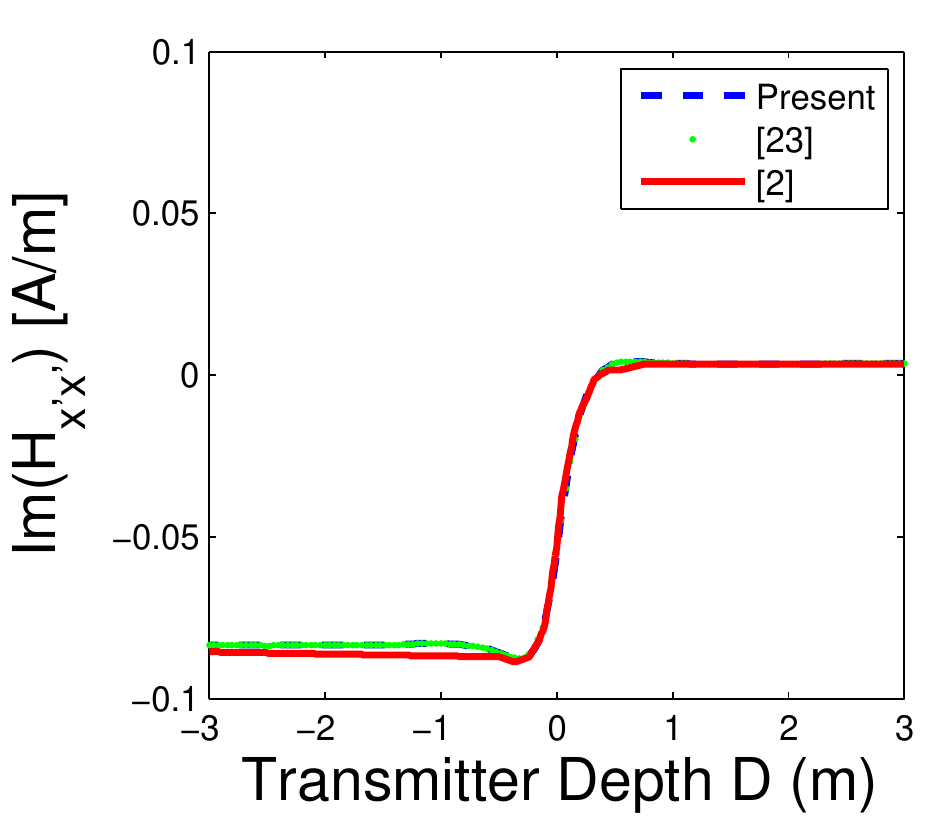}}

\subfloat[\label{Sofia4a3R}]{\includegraphics[width=2.7in]{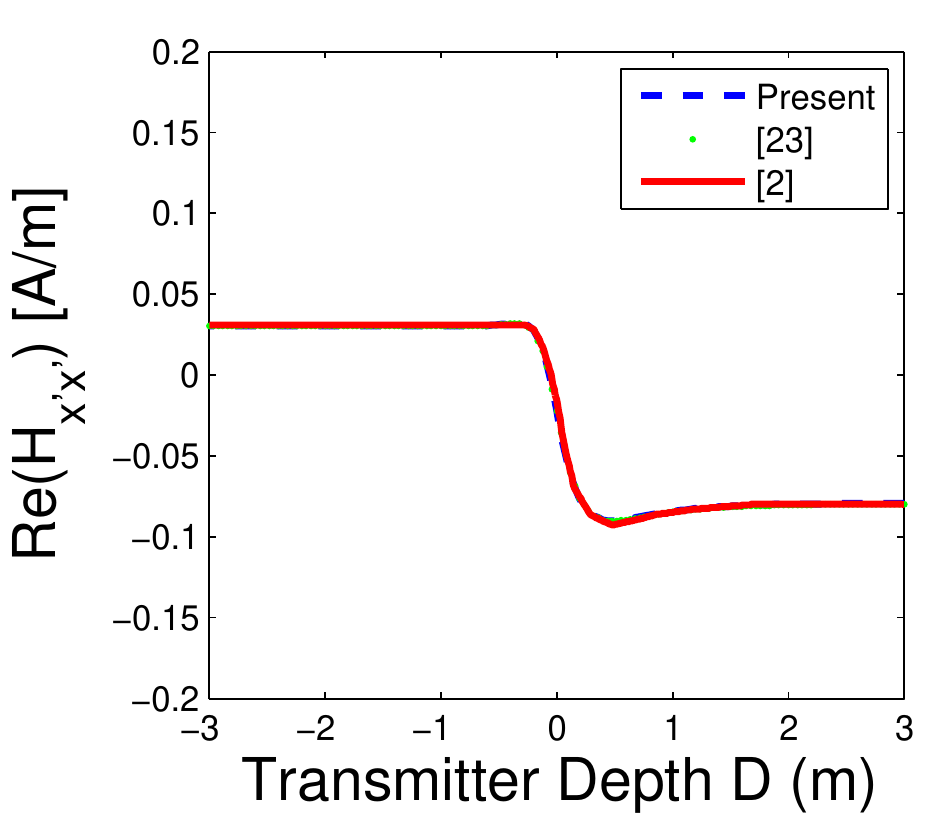}}
\subfloat[\label{Sofia4a3I}]{\includegraphics[width=2.7in]{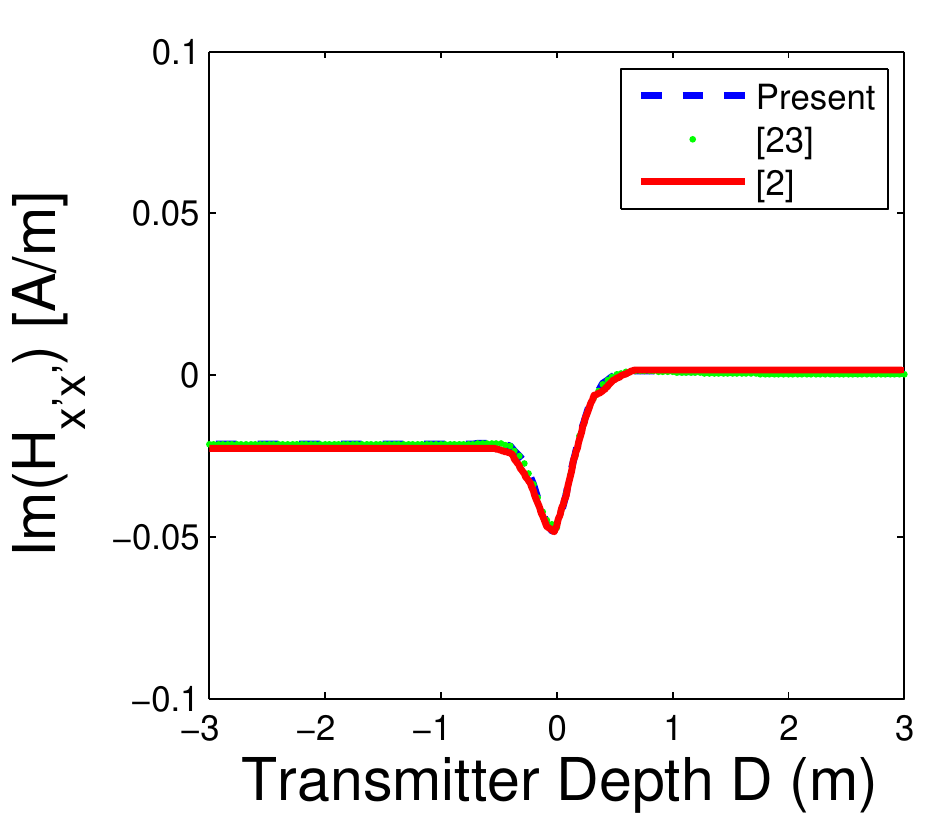}}
\caption{\small Comparison of computed magnetic field $H_{x'x'}$ against results from Figure 4 of \cite{sofia}.}
\label{SofiaFig4Hxx}
    \end{figure}

\begin{figure}[H]
\centering
\subfloat[\label{Sofia4b1R}]{\includegraphics[width=2.7in]{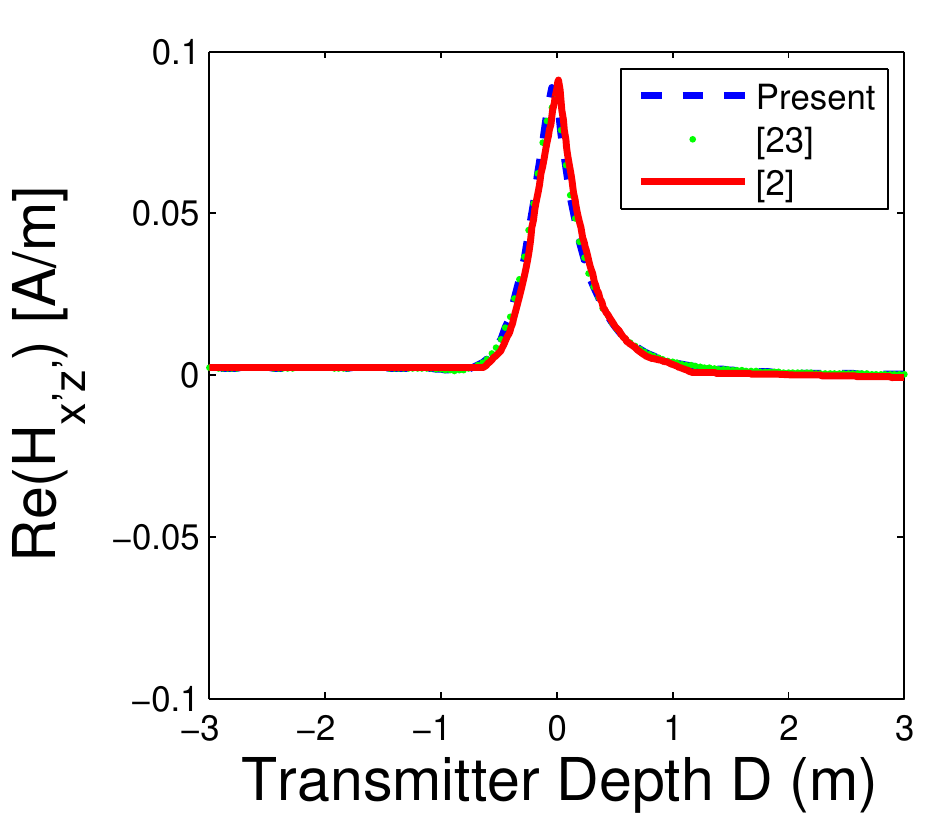}}
\subfloat[\label{Sofia4b1I}]{\includegraphics[width=2.7in]{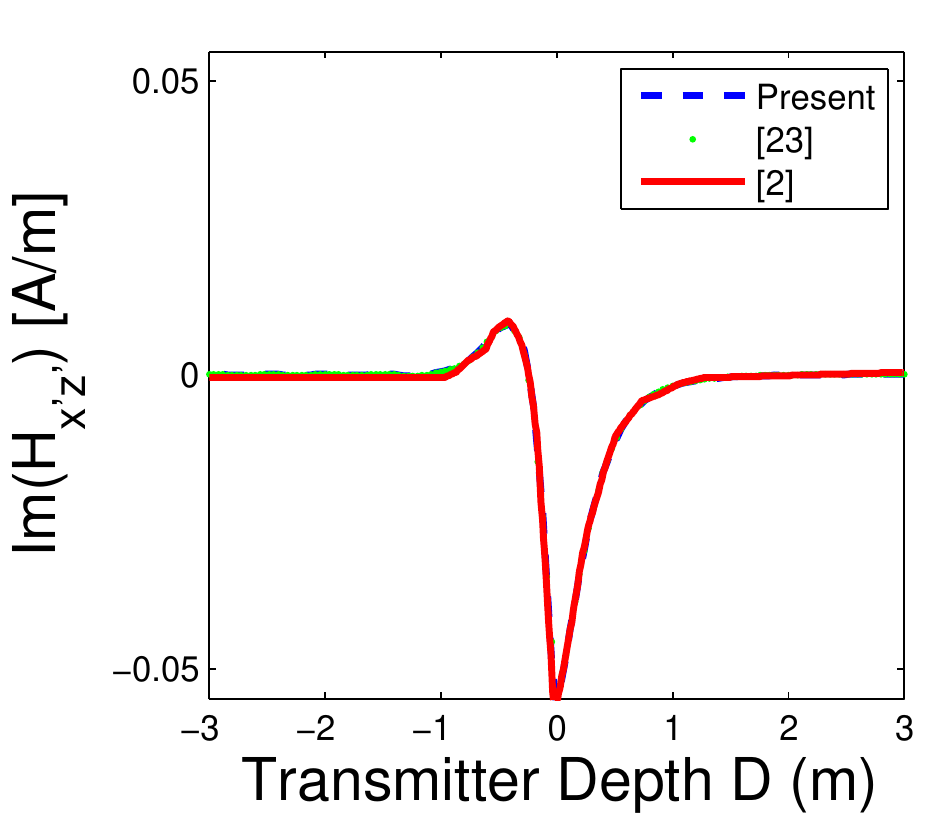}}

\subfloat[\label{Sofia4b2R}]{\includegraphics[width=2.7in]{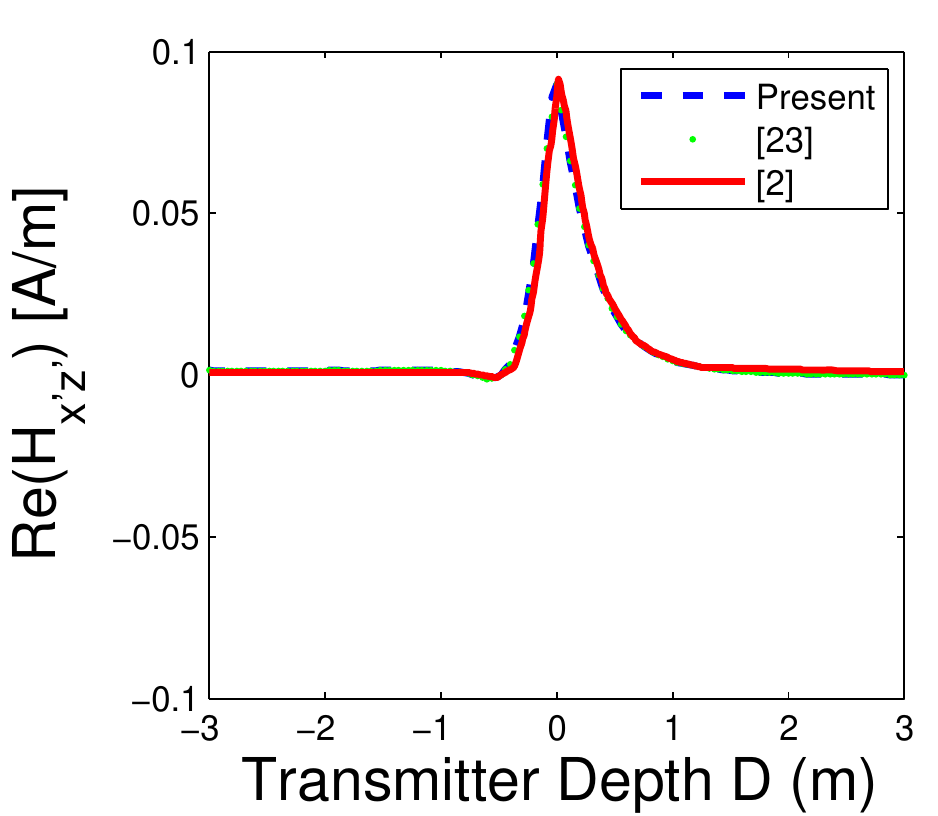}}
\subfloat[\label{Sofia4b2I}]{\includegraphics[width=2.7in]{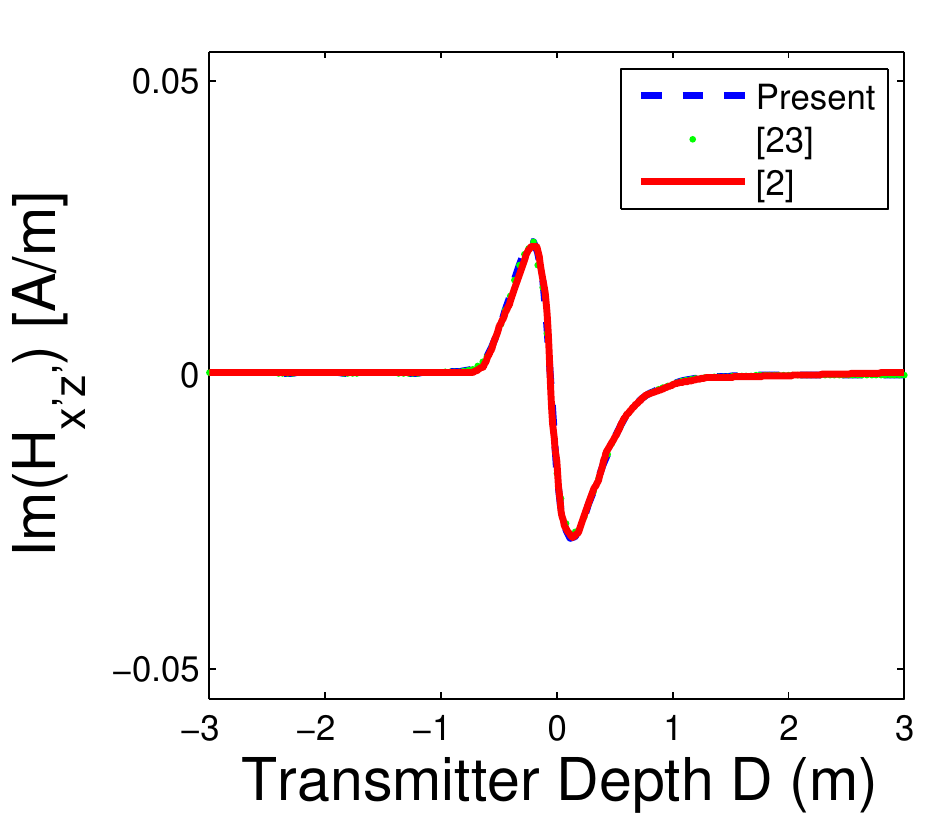}}

\subfloat[\label{Sofia4b3R}]{\includegraphics[width=2.7in]{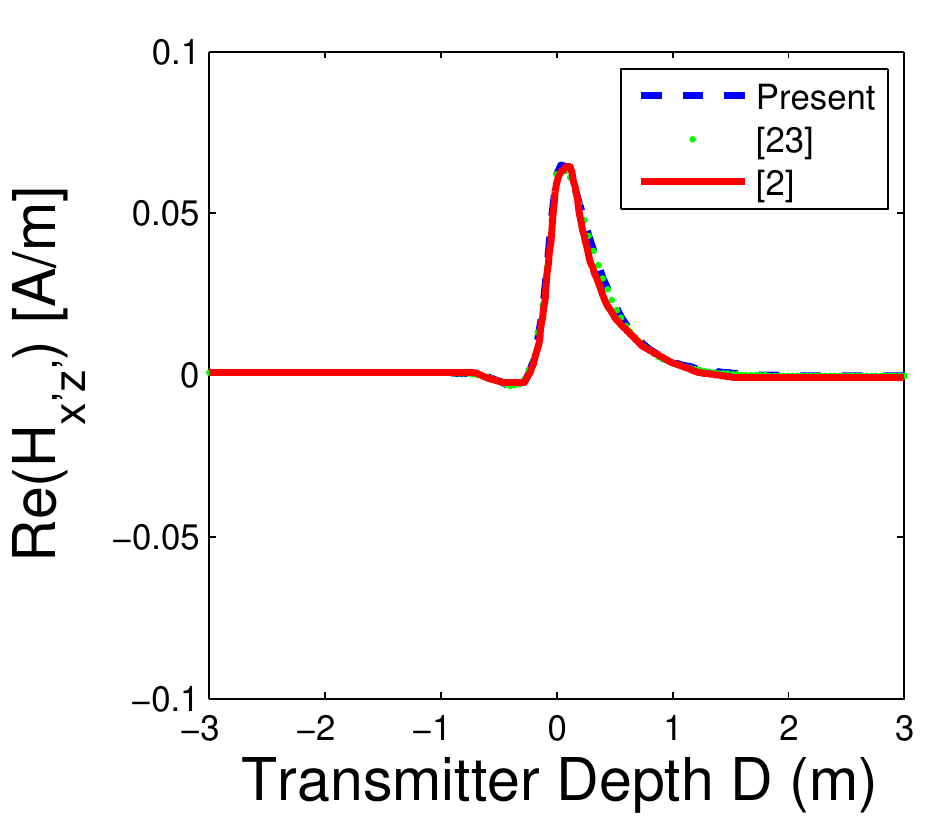}}
\subfloat[\label{Sofia4b3I}]{\includegraphics[width=2.7in]{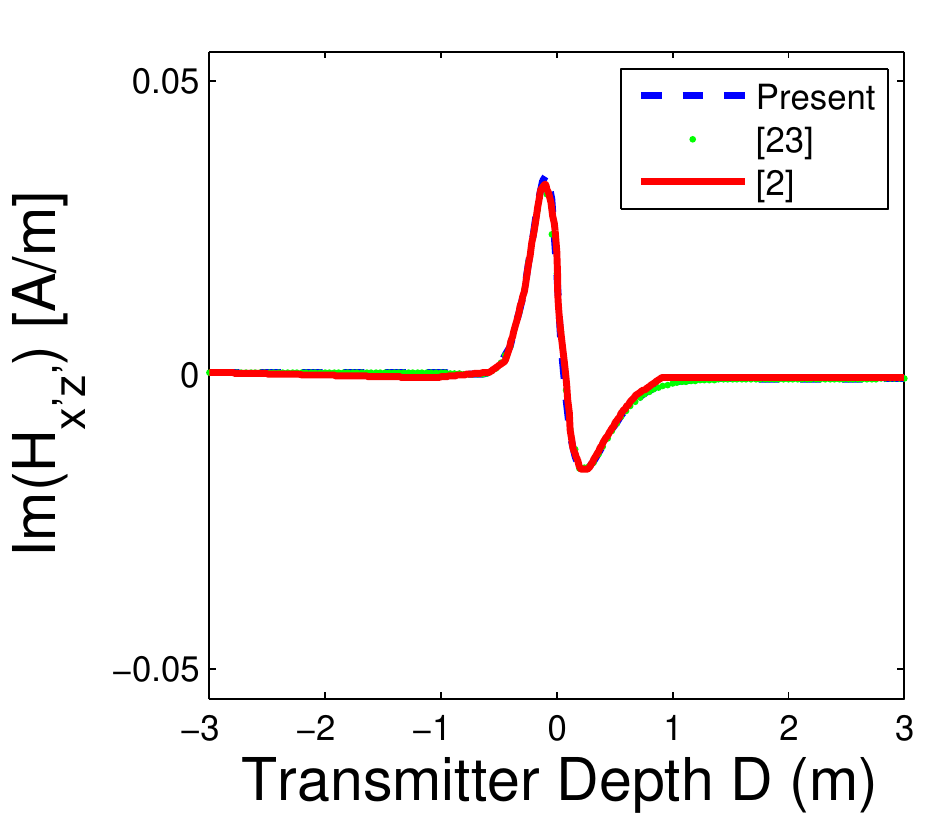}}
\caption{\small Comparison of computed magnetic field $H_{x'z'}$ against results from Figure 4 of \cite{sofia}.}
\label{SofiaFig4Hxz}
    \end{figure}

\begin{figure}[H]
\centering
\subfloat[\label{Sofia4d1R}]{\includegraphics[width=2.7in]{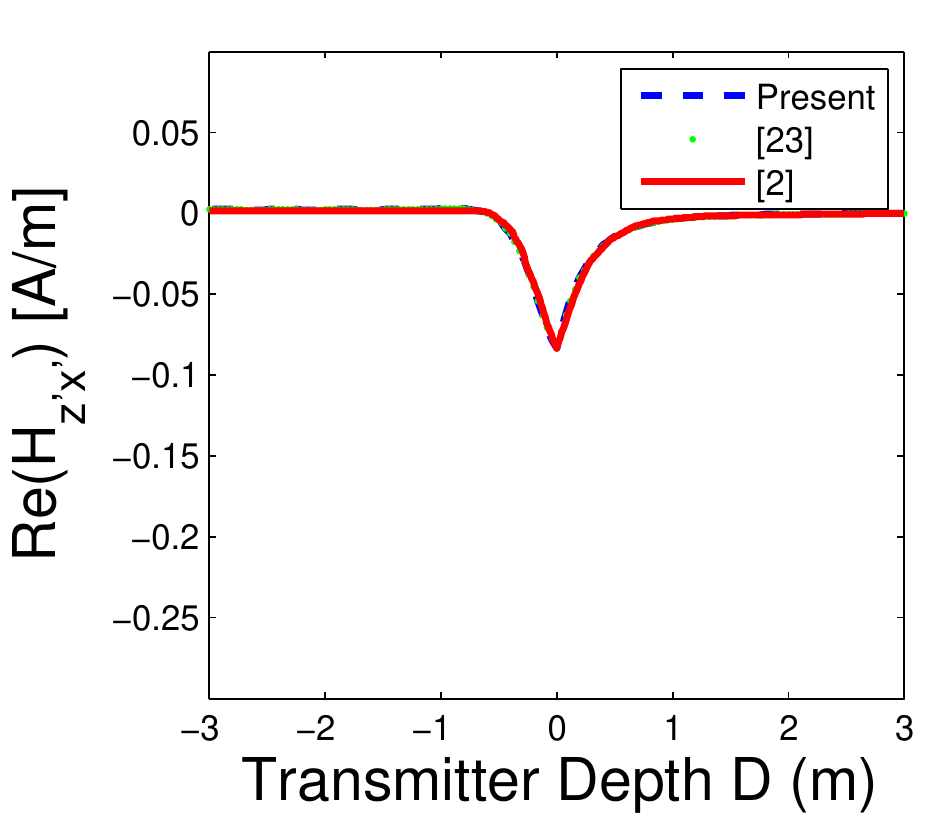}}
\subfloat[\label{Sofia4d1I}]{\includegraphics[width=2.7in]{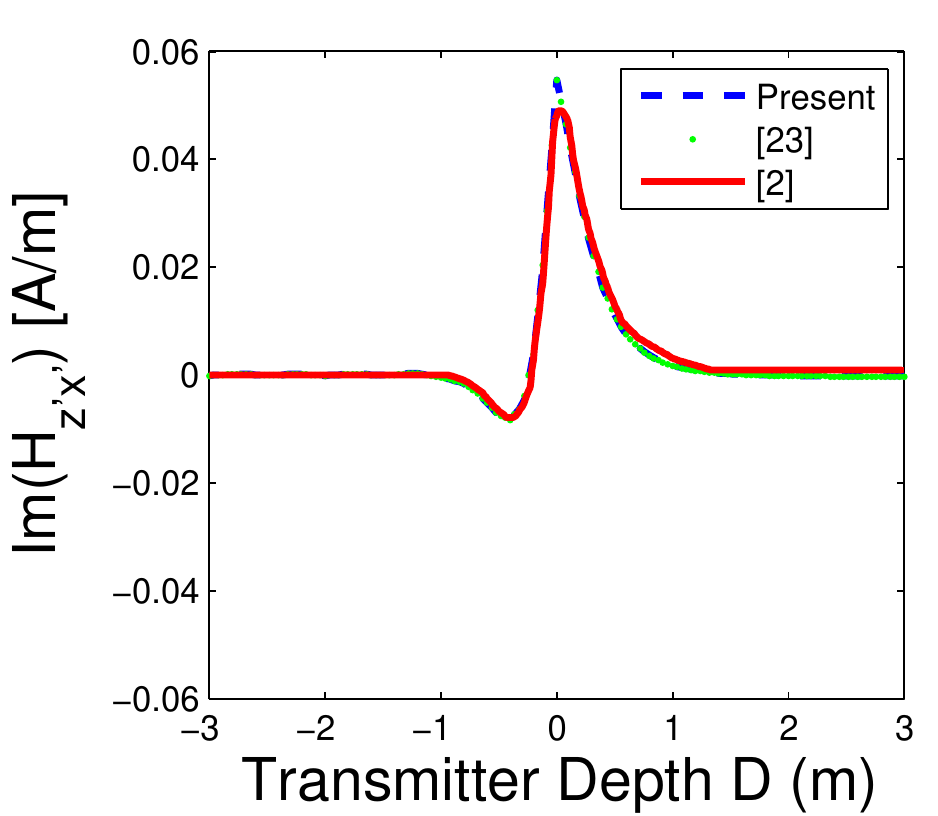}}

\subfloat[\label{Sofia4d2R}]{\includegraphics[width=2.7in]{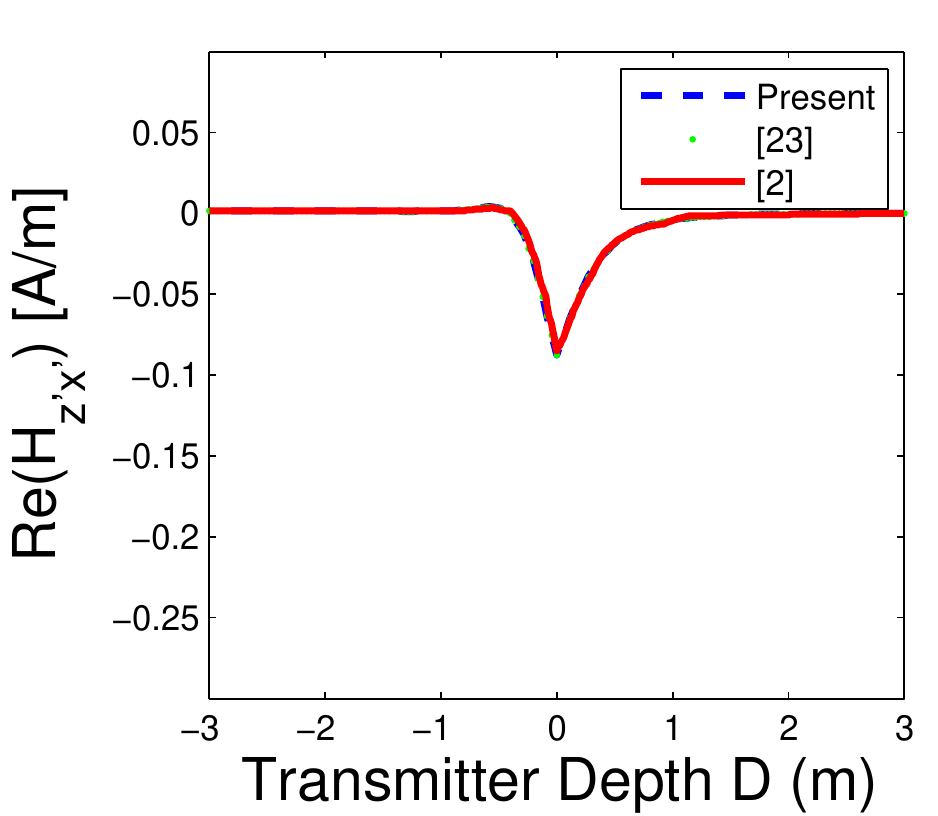}}
\subfloat[\label{Sofia4d2I}]{\includegraphics[width=2.7in]{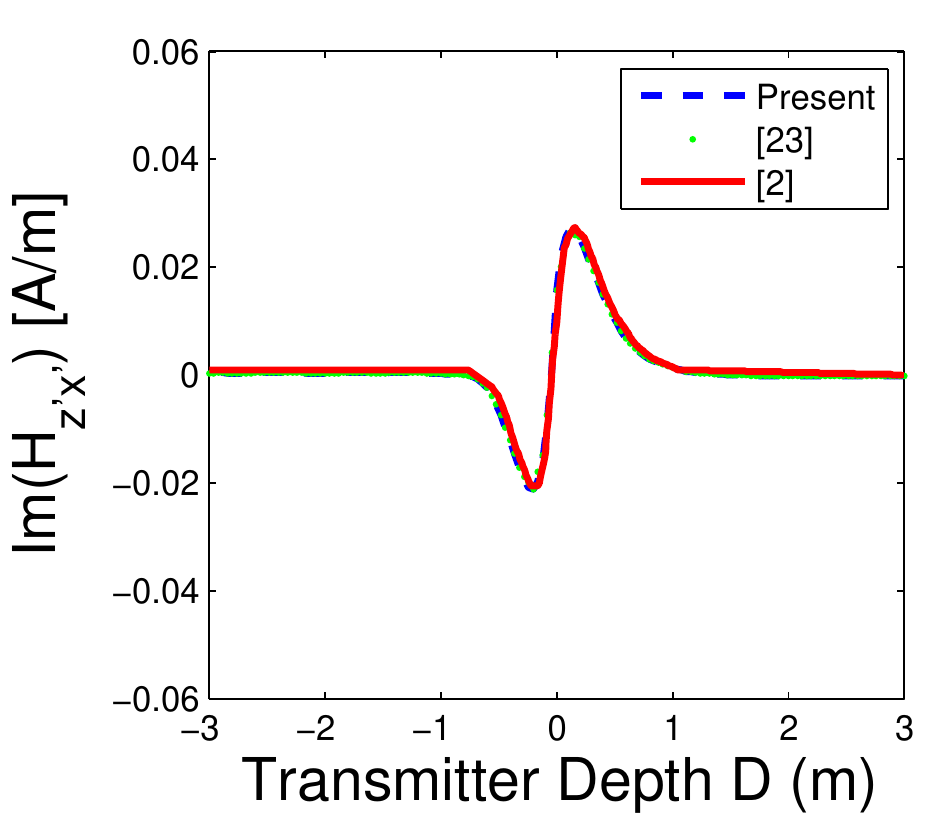}}

\subfloat[\label{Sofia4d3R}]{\includegraphics[width=2.7in]{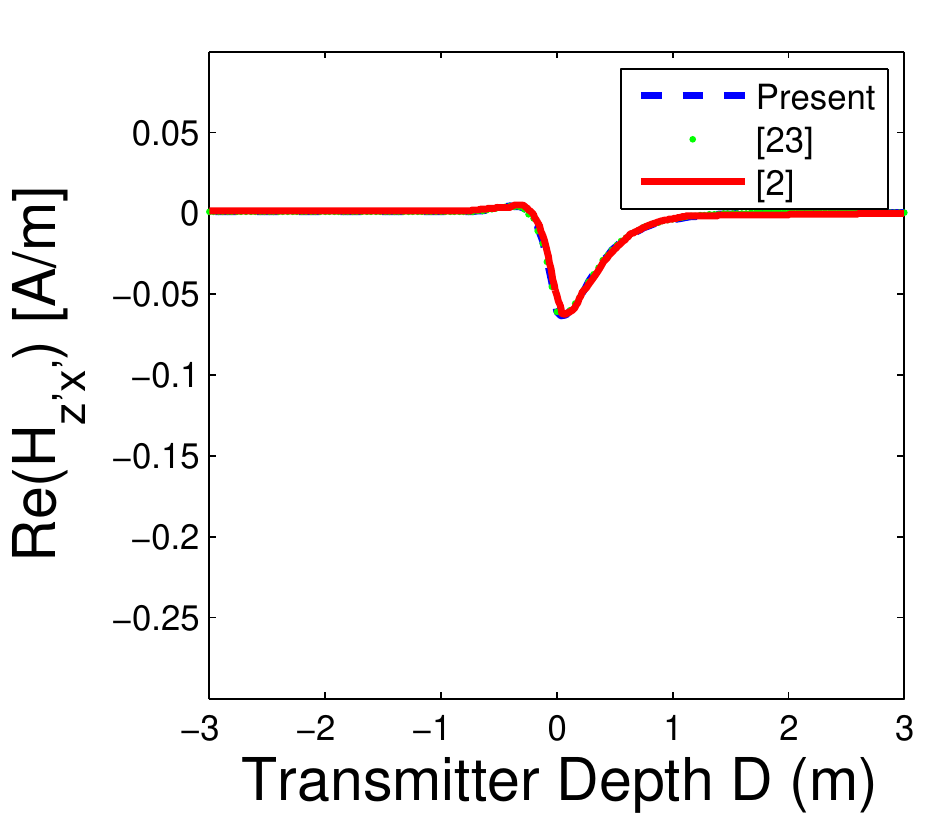}}
\subfloat[\label{Sofia4d3I}]{\includegraphics[width=2.7in]{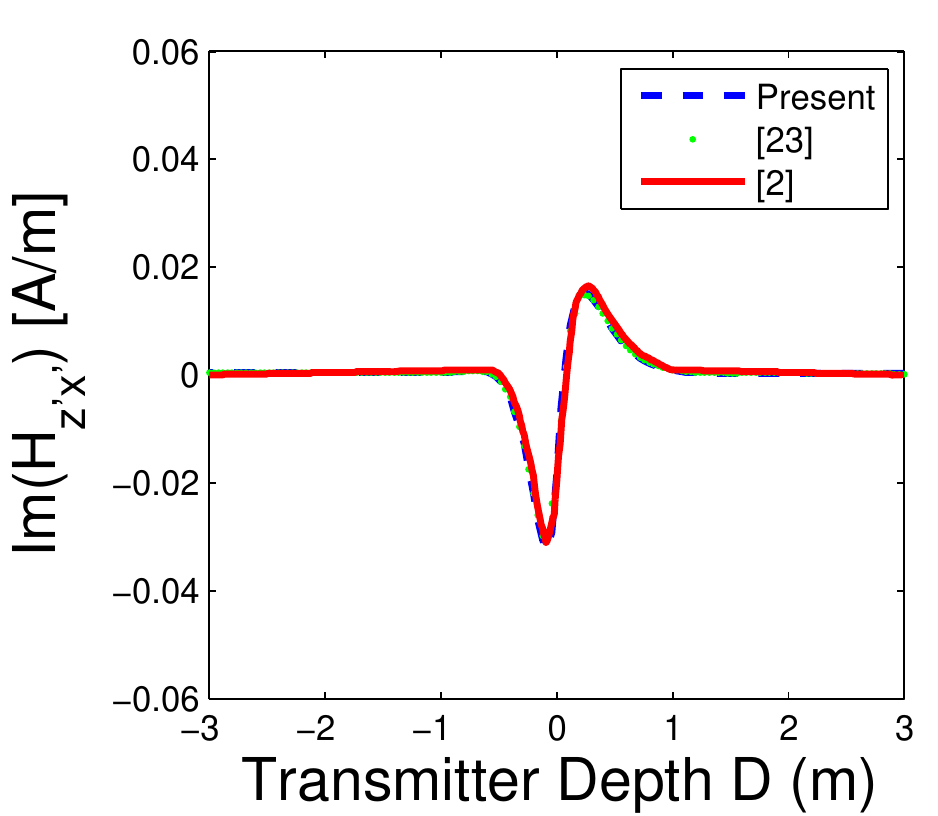}}
\caption{\small Comparison of computed magnetic field $H_{z'x'}$ against results from Figure 4 of \cite{sofia}.}
\label{SofiaFig4Hzx}
    \end{figure}

\begin{figure}[H]
\centering
\subfloat[\label{Sofia4e1R}]{\includegraphics[width=2.7in]{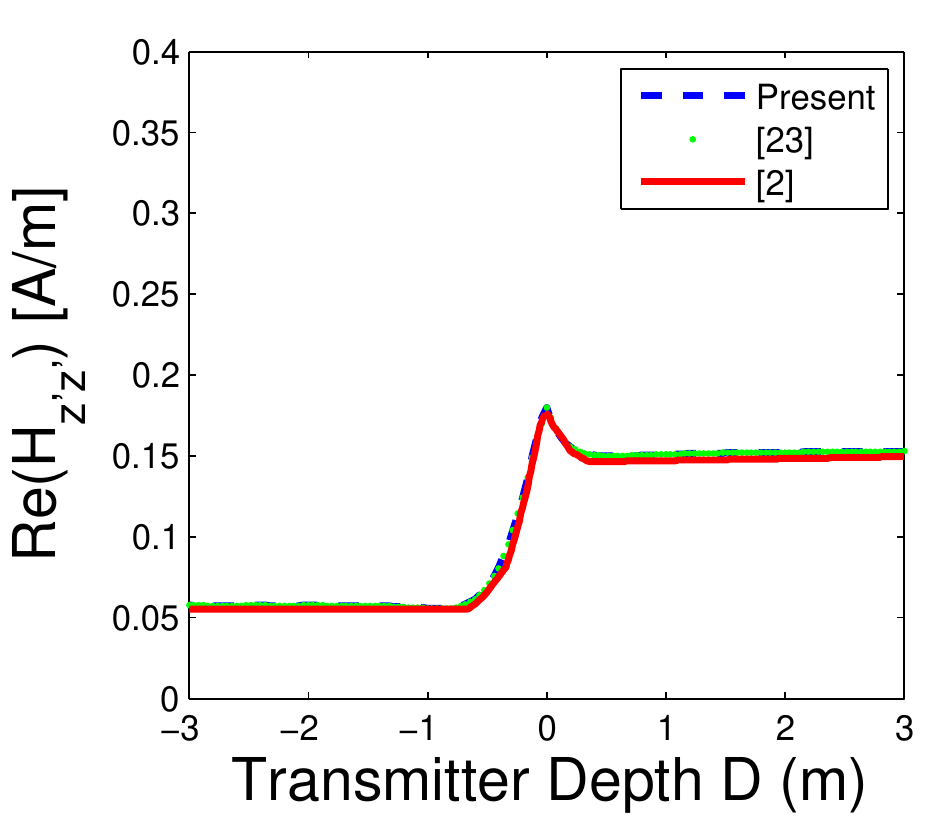}}
\subfloat[\label{Sofia4e1I}]{\includegraphics[width=2.7in]{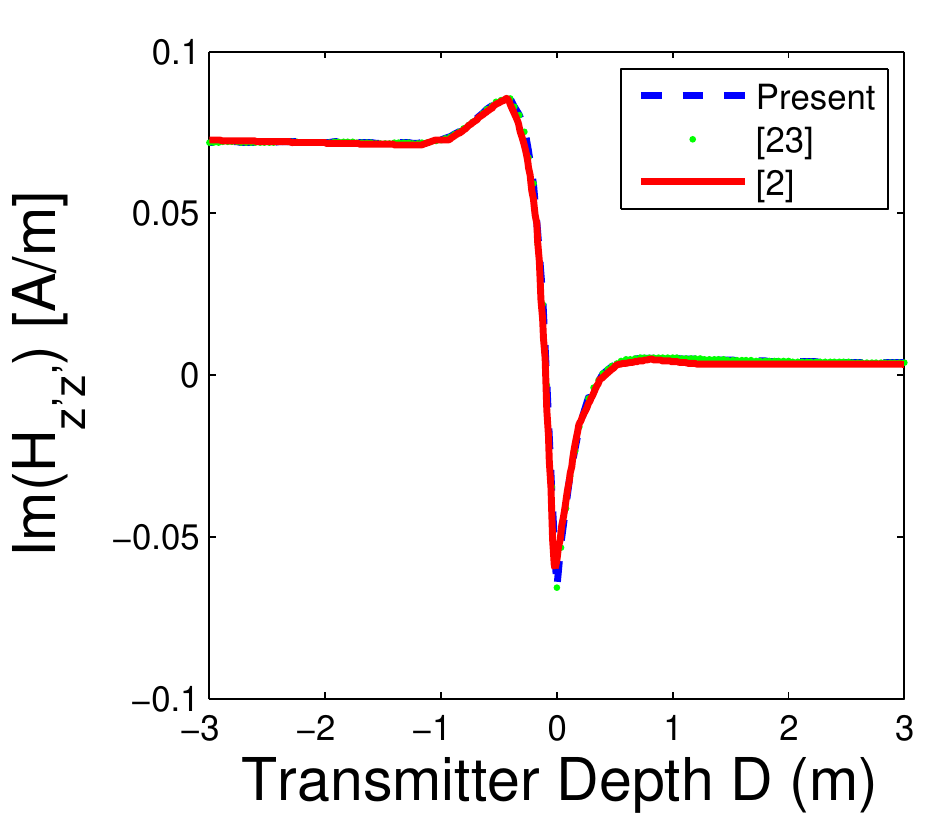}}

\subfloat[\label{Sofia4e2R}]{\includegraphics[width=2.7in]{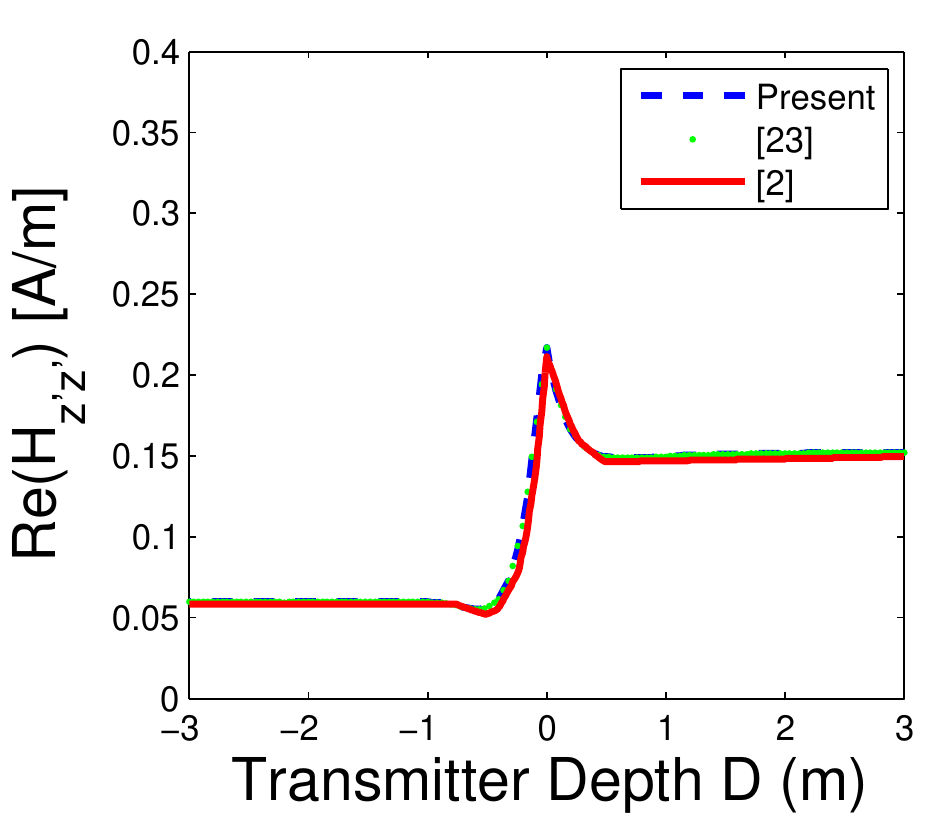}}
\subfloat[\label{Sofia4e2I}]{\includegraphics[width=2.7in]{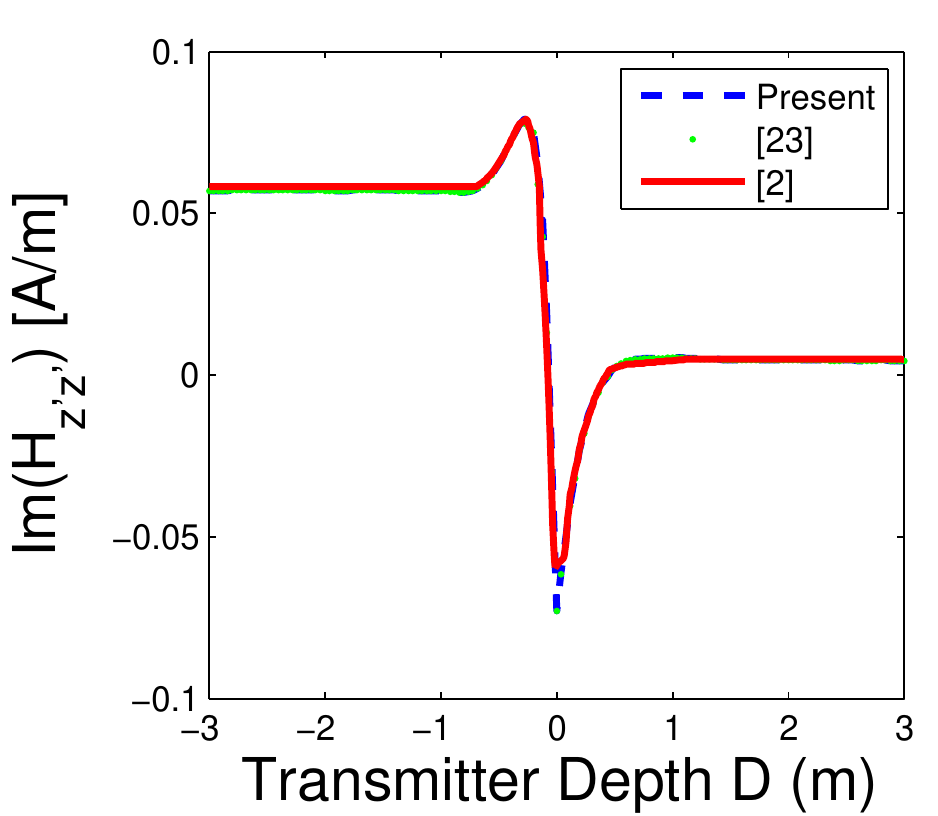}}

\subfloat[\label{Sofia4e3R}]{\includegraphics[width=2.7in]{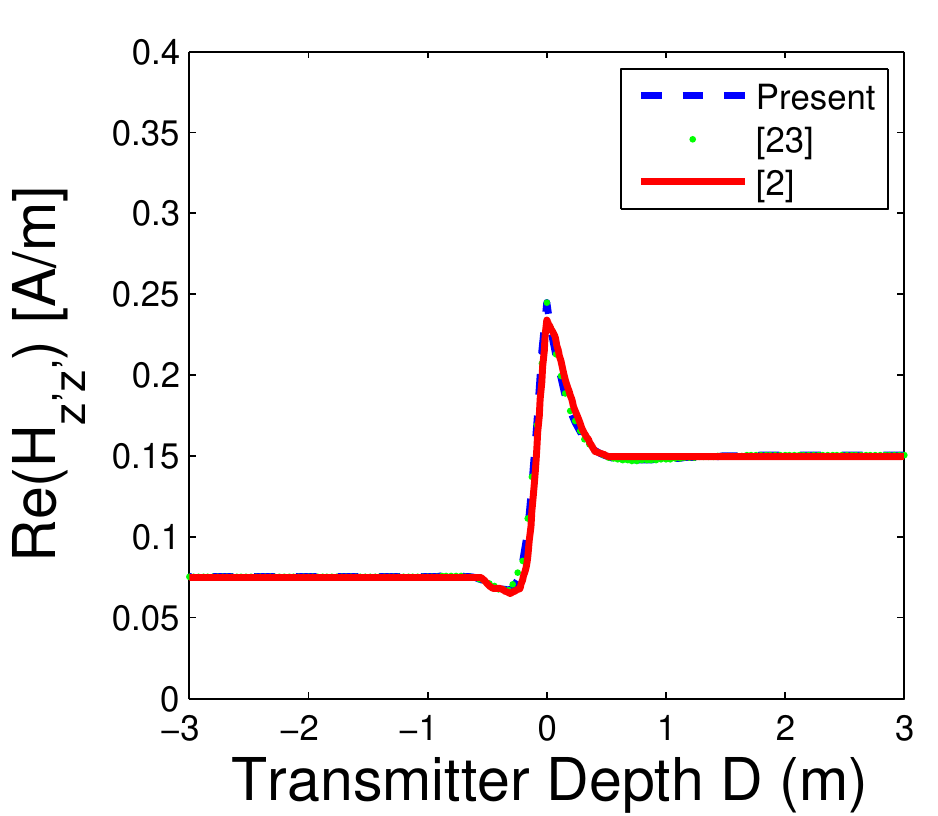}}
\subfloat[\label{Sofia4e3I}]{\includegraphics[width=2.7in]{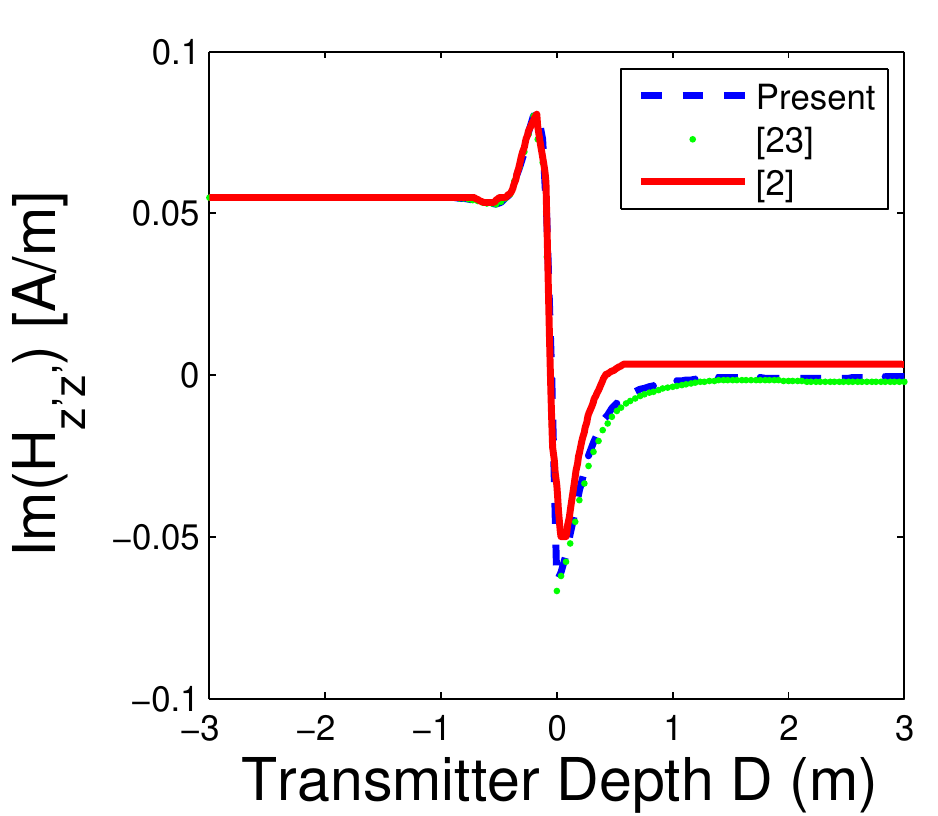}}
\caption{\small Comparison of computed magnetic field $H_{z'z'}$ against results from Figure 4 of \cite{sofia}.}
\label{SofiaFig4Hzz}
    \end{figure}

\subsection{Planar Antenna Above Doubly-Anisotropic Isoimpedance Substrates}
\begin{figure}[H]
\centering
\subfloat[\label{Case1}]{\includegraphics[width=2in]{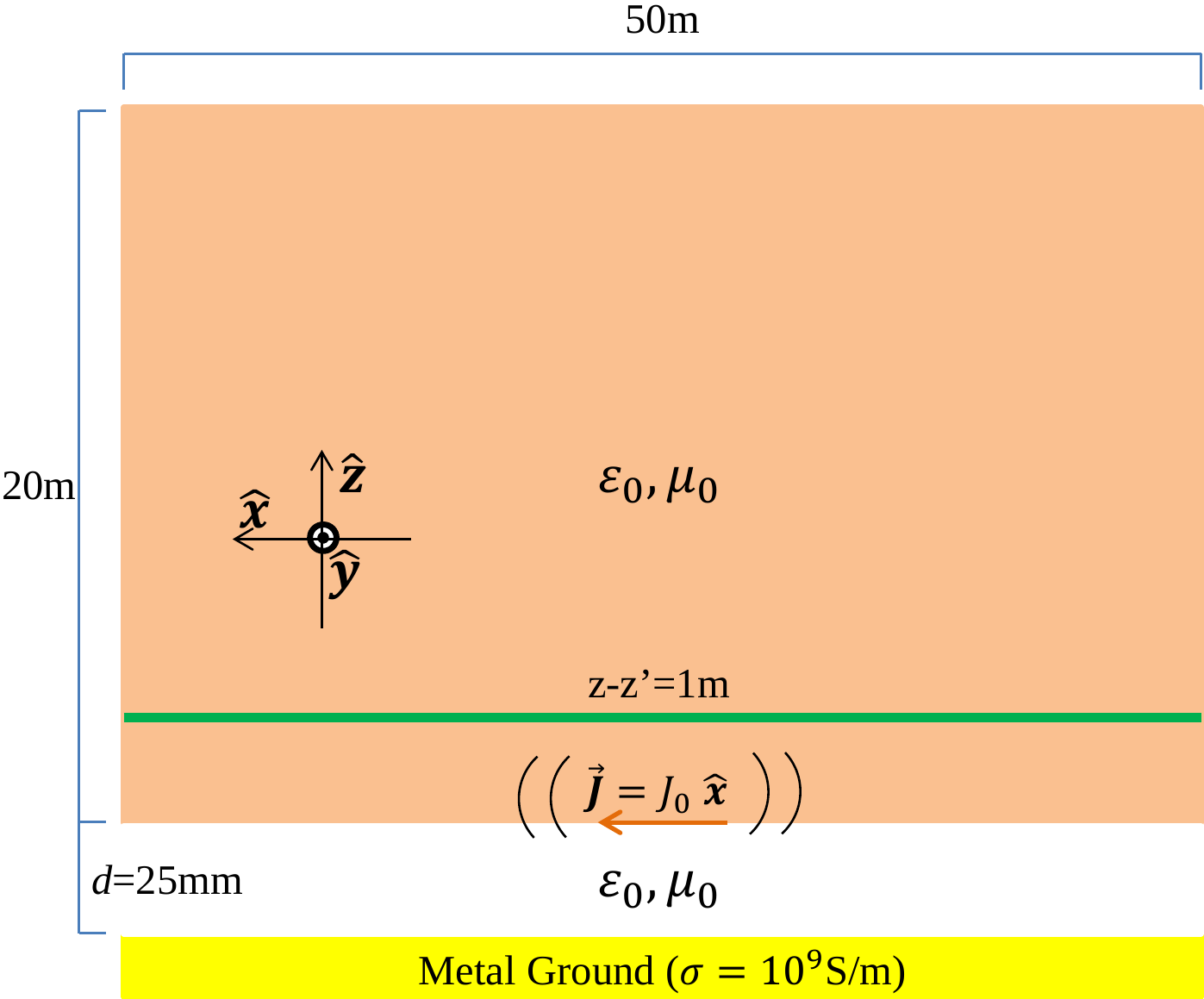}}
\subfloat[\label{Case2}]{\includegraphics[width=2in]{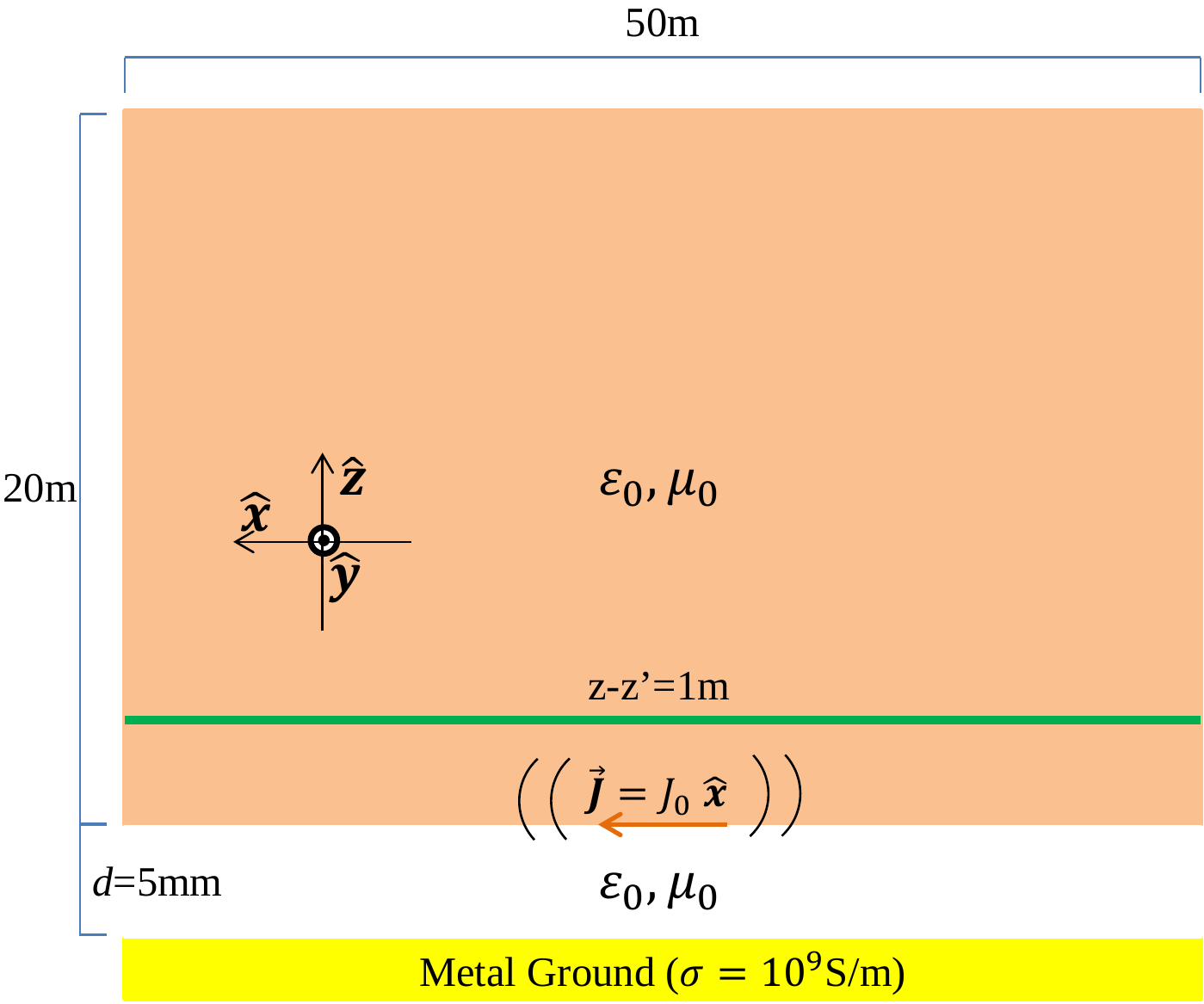}}
\subfloat[\label{Case3}]{\includegraphics[width=2.1in]{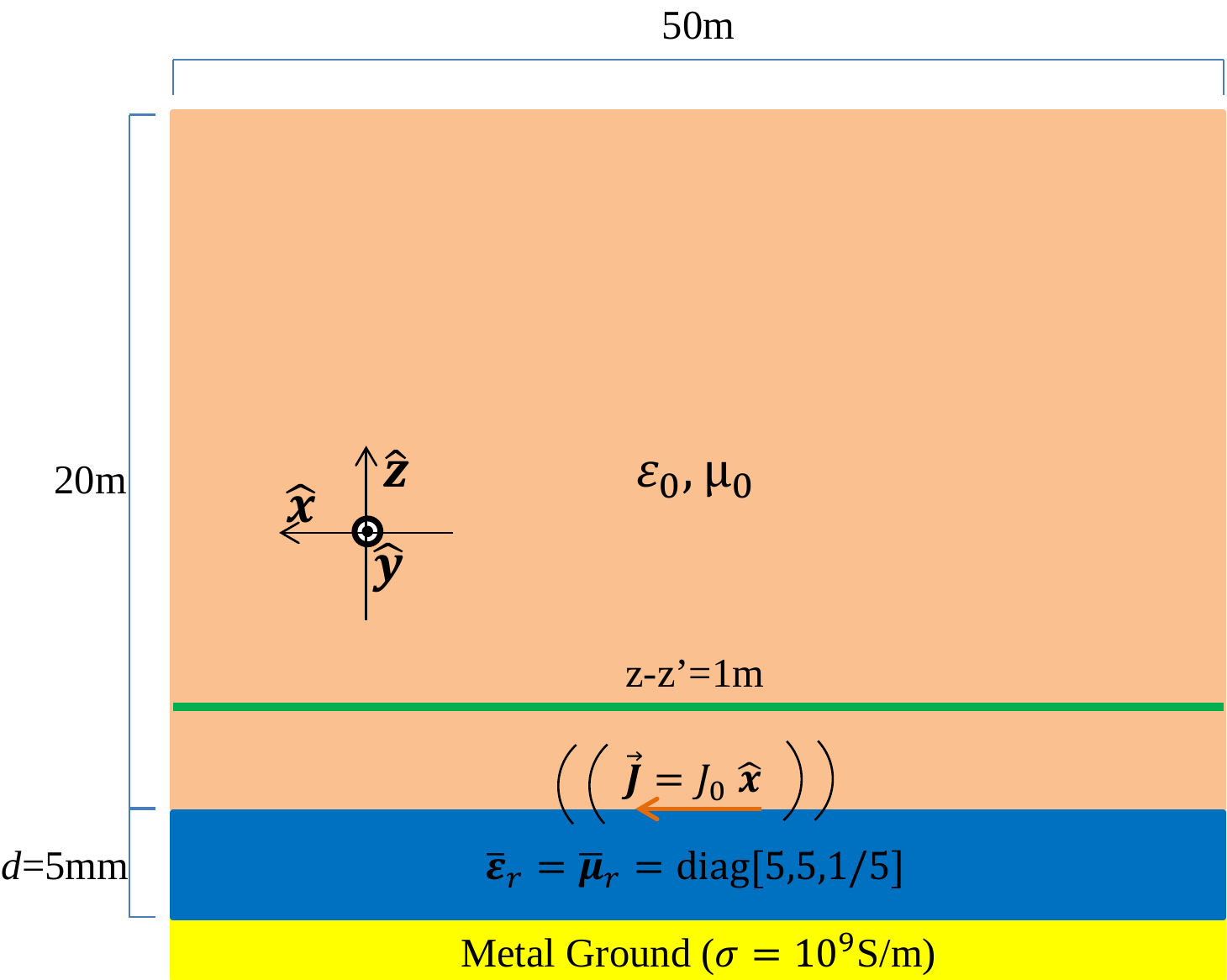}}
\caption{\small \label{CasePic}Schematic illustration of the three scenarios simulated. The Hertzian electric dipole is always oriented in the $+x$ direction, radiates at $f=13.56$MHz, and is located a distance $d$ above the ground plane, which has a conductivity $\sigma=10^9$ S/m. The light brown region indicates the region of observation in free space for the exhibited $xz$ plane ($y-y'=0$) electric field distribution plots in Figures \ref{Iso1a}, \ref{Iso1c}, and \ref{Iso1e}, while the region of observation for the magnetic field distribution plots in Figures \ref{Iso1b}, \ref{Iso1d}, and \ref{Iso1f} is obtained upon rotating this light brown-colored plane by ninety degrees about the $z$ axis, yielding the $yz$ plane ($x-x'=0$). Finally, the constant-$z$ plane indicated by the green line in each sub-figure of Figure \ref{CasePic} indicates the location of the $xy$ plane cut on which $|E_z|$ and $|H_z|$ are plotted in Figure \ref{Iso2}. Note that, contrary to the situation suggested in Figure \ref{CasePic}, the ground plane is assumed infinite in its lateral extent while the observation plane is laterally bounded.}
\label{Cases}
\end{figure}

Next, we illustrate the application of the proposed algorithm to the modeling of planar radiators on top of isoimpedance anisotropic substrates backed by metallic ground planes. Isoimpedance substrates are substrates utilized to minimize the antenna profile by reducing substrate thickness. Conventionally, to reduce the strong field cancelation effect caused by the presence of a ground plane, substrates made of (for example) dielectric or ferrite material are used \cite{teixeira1,balanis1}. Such conventional substrates typically exhibit various disadvantages such as high ohmic loss, unwanted surface waves (and hence reduced radiation efficiency and realized antenna gain), reduced bandwidth \cite{balanis1,teixeira1}, and the need for large thickness to yield a useful radiation resistance. On the other hand, isoimpedance substrates can facilitate a miniaturized longitudinal profile by mimicking the effect of a thicker substrate \cite{teixeira1,pendry1}. Furthermore, since isoimpedance media are inherently impedance-matched to free space for all wave incidence angles \cite{teixeiraJEWA}, they do not support surface waves \cite{teixeira1}. The problem under consideration is illustrated in Figures \ref{Case1},\ref{Case2}, and \ref{Case3}, which show a lateral view of the geometry.

\begin{figure}[H]
\centering
\subfloat[\label{Iso1a}]{\includegraphics[width=2.5in]{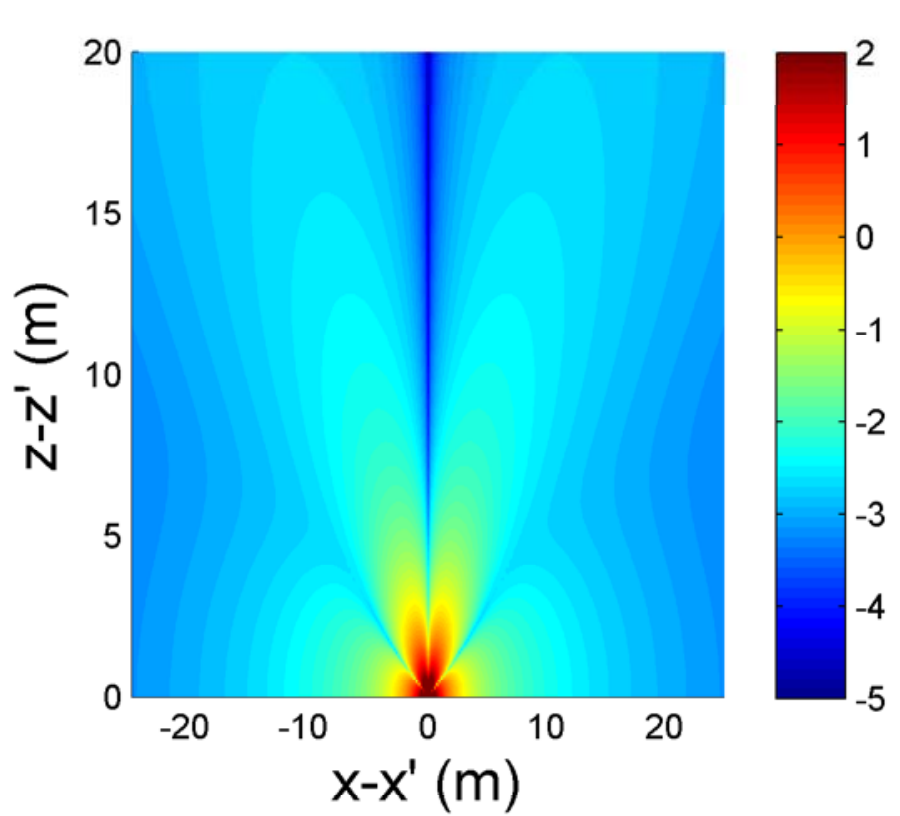}}
\subfloat[\label{Iso1b}]{\includegraphics[width=2.5in]{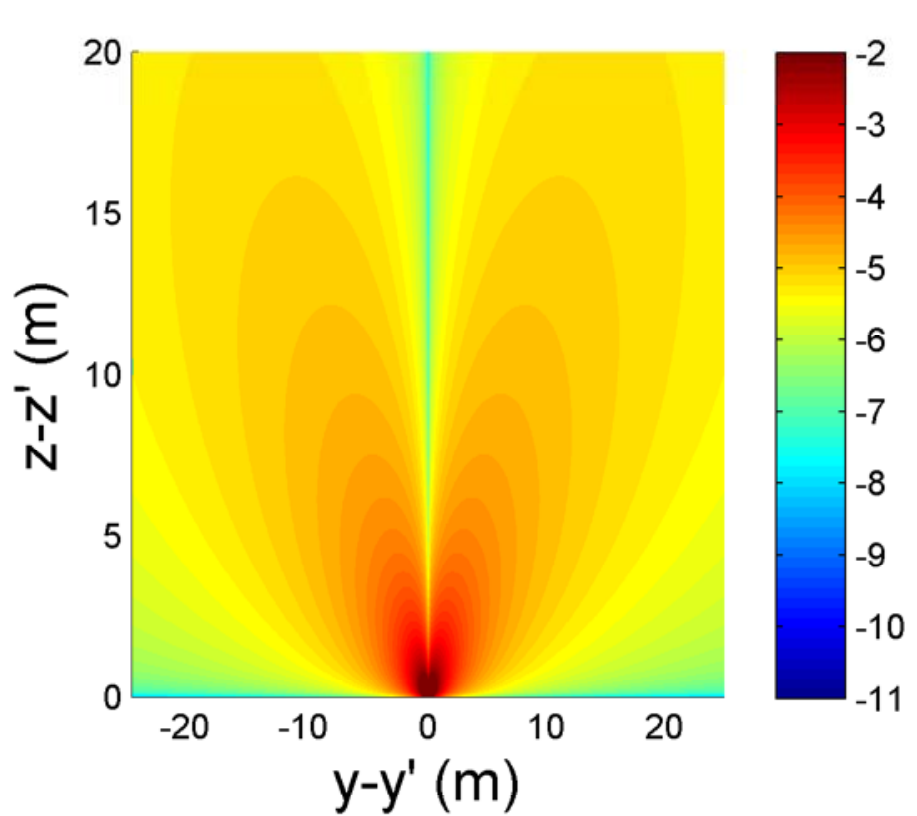}}

\subfloat[\label{Iso1c}]{\includegraphics[width=2.5in]{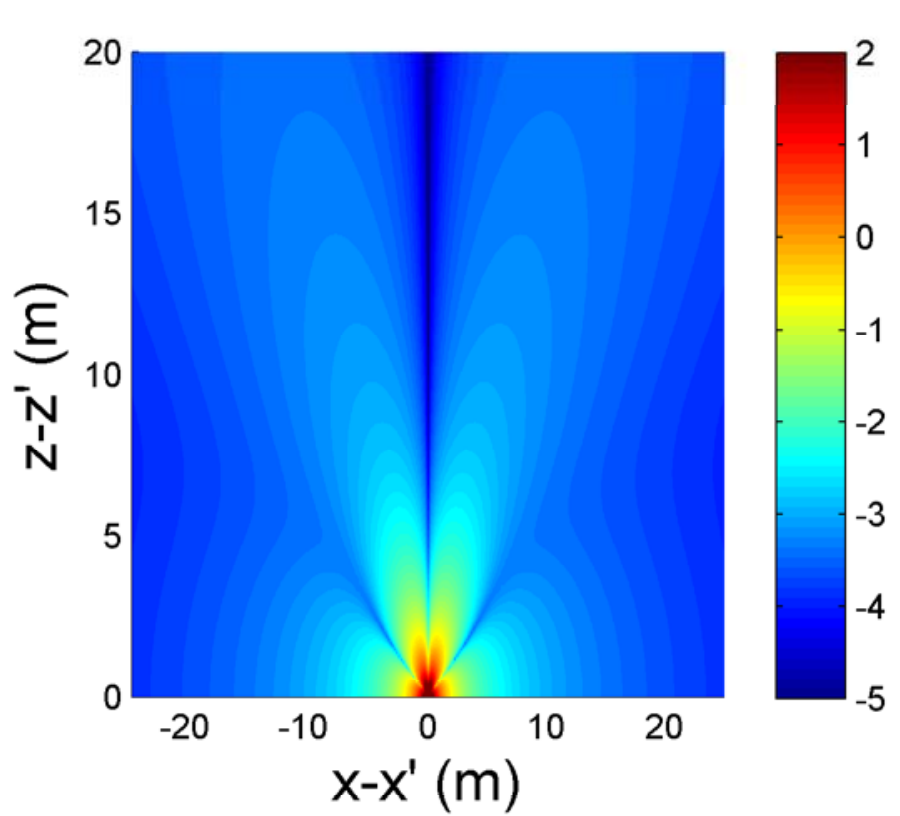}}
\subfloat[\label{Iso1d}]{\includegraphics[width=2.5in]{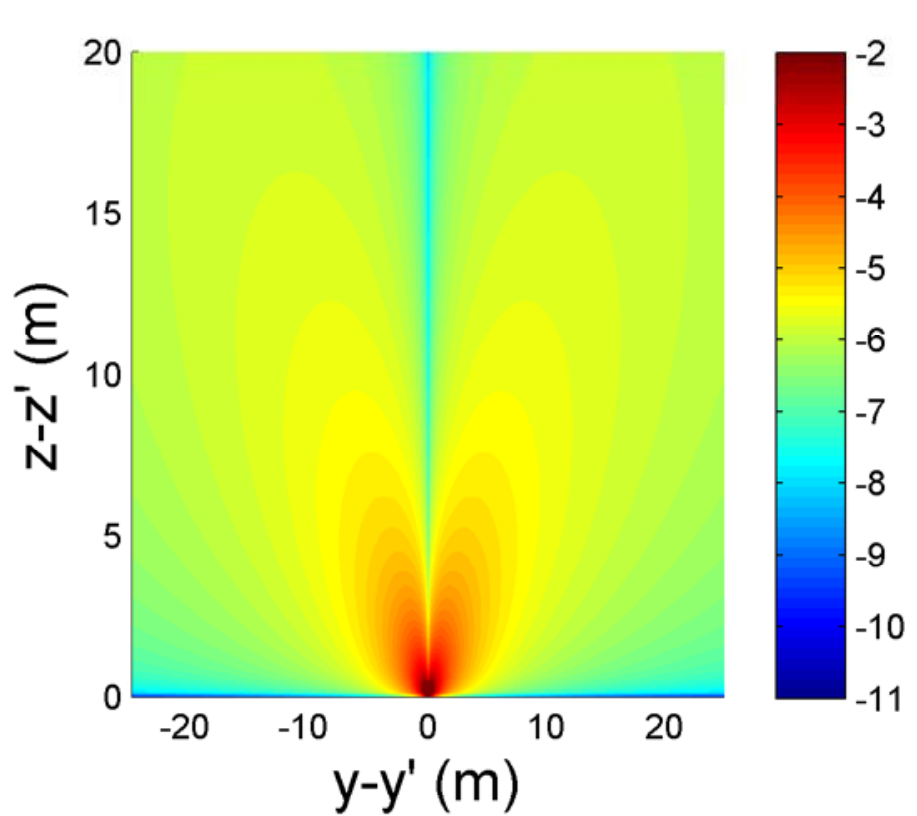}}

\subfloat[\label{Iso1e}]{\includegraphics[width=2.5in]{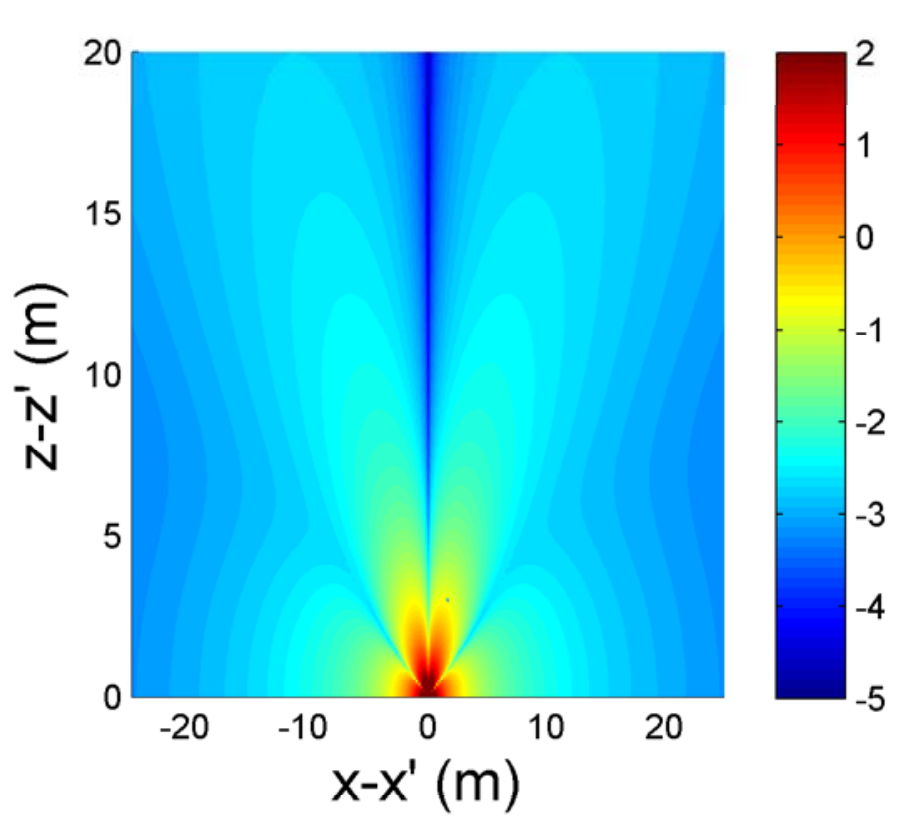}}
\subfloat[\label{Iso1f}]{\includegraphics[width=2.5in]{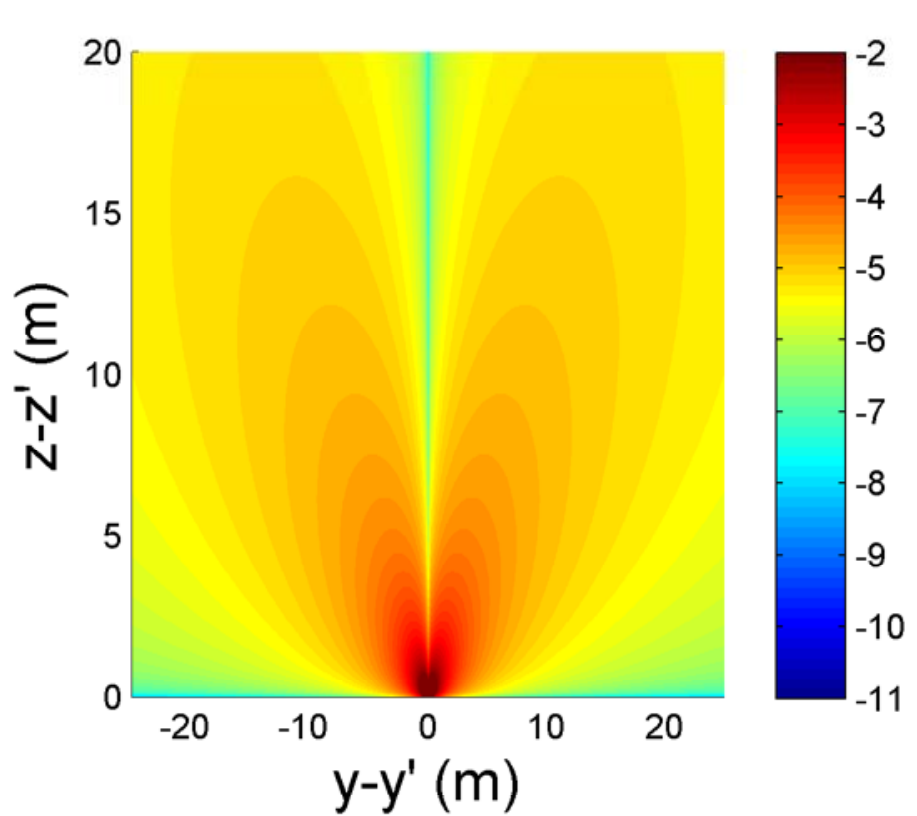}}
\caption{\small Electric field $|E_z|$ distribution (first column) and magnetic field $|H_z|$ distribution (right column) due to a Hertzian electric dipole located at $(0,0,d)$m.}
\label{Iso1}
    \end{figure}

\begin{figure}[H]
\centering
\subfloat[\label{Iso2a}]{\includegraphics[width=2.5in]{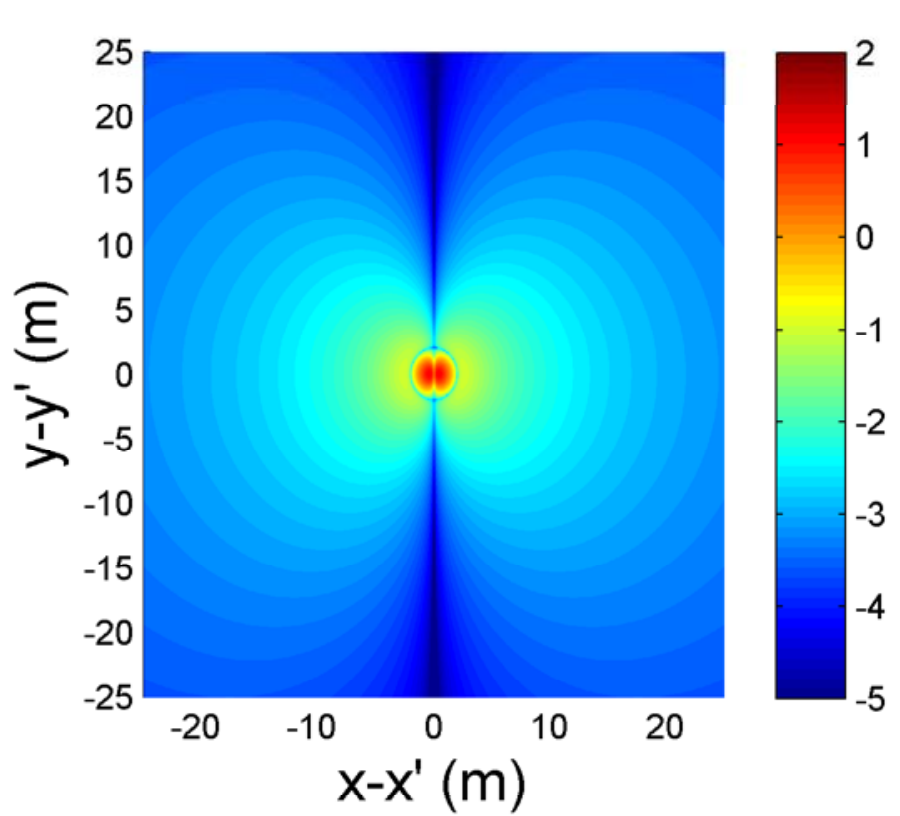}}
\subfloat[\label{Iso2b}]{\includegraphics[width=2.5in]{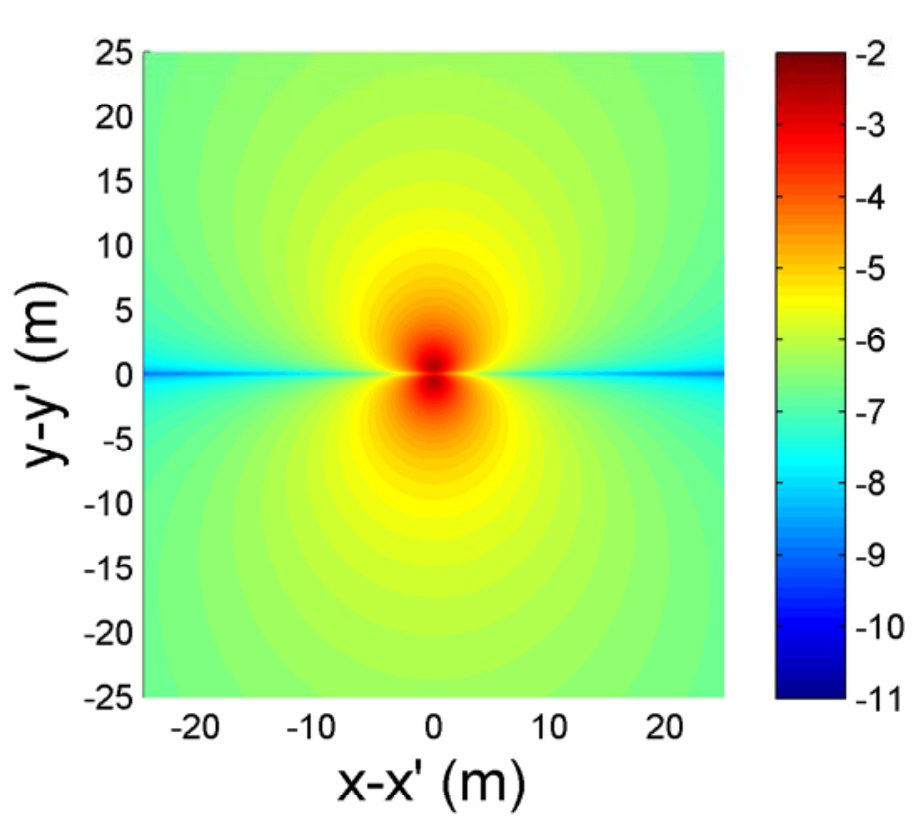}}

\subfloat[\label{Iso2c}]{\includegraphics[width=2.5in]{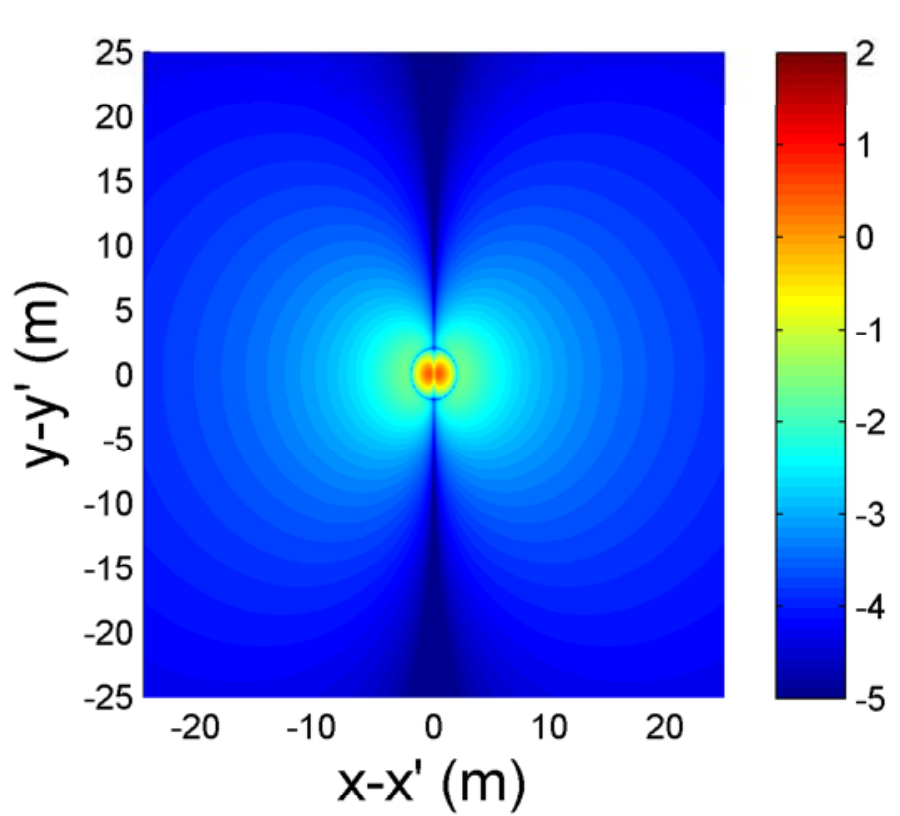}}
\subfloat[\label{Iso2d}]{\includegraphics[width=2.5in]{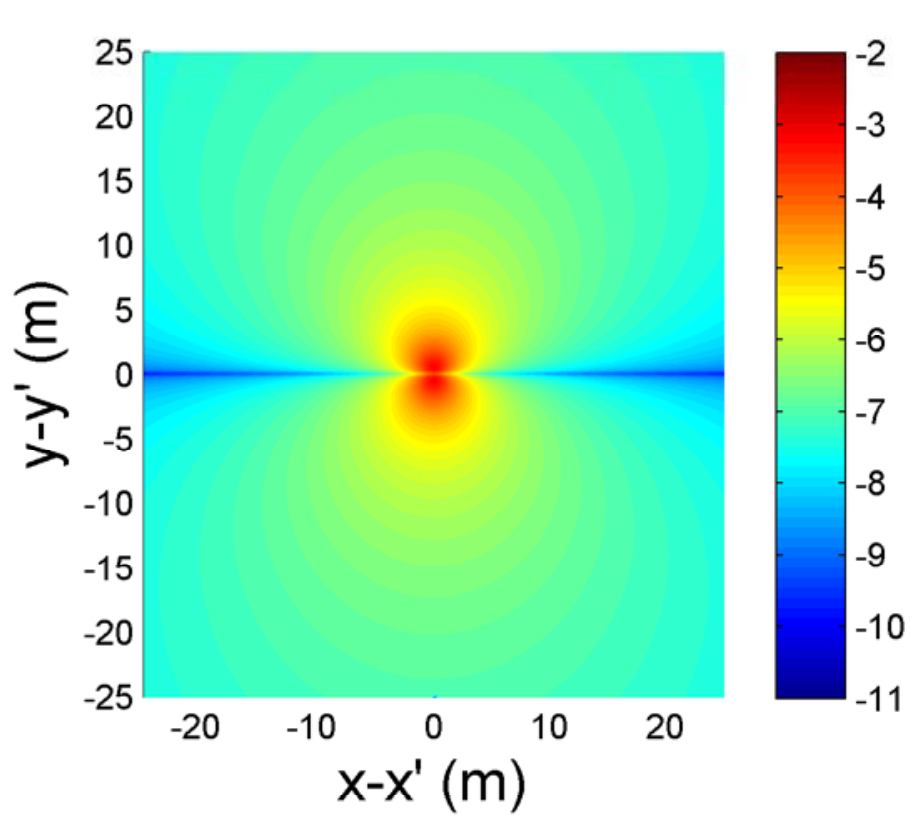}}

\subfloat[\label{Iso2e}]{\includegraphics[width=2.5in]{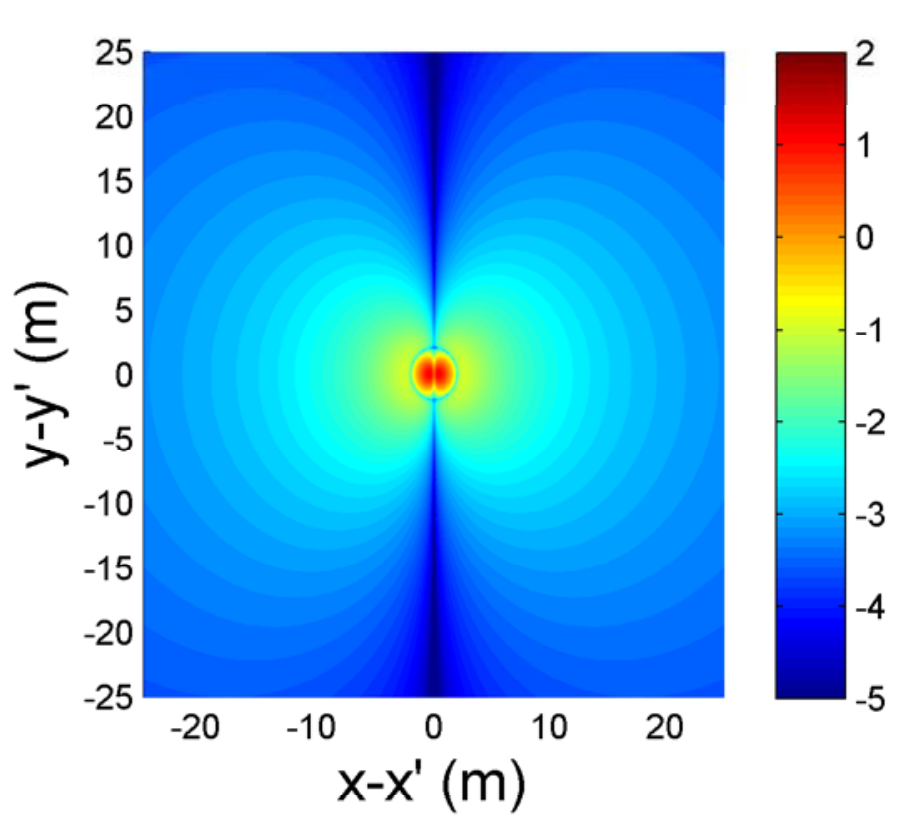}}
\subfloat[\label{Iso2f}]{\includegraphics[width=2.5in]{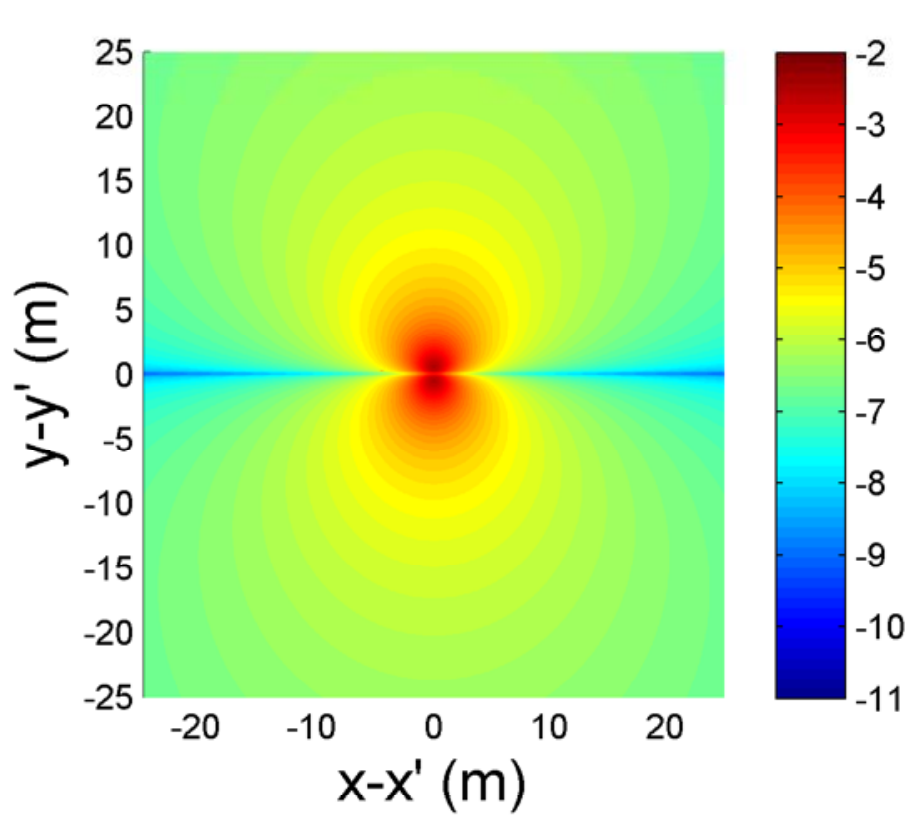}}
\caption{\small Each row of plots corresponds to the same respective environment scenarios as Figure \ref{Iso1}, except both $|E_z|$ and $|H_z|$ are plotted on an $xy$ plane cut (see Figure \ref{CasePic}).}
\label{Iso2}
    \end{figure}

The field distributions are presented in Figures \ref{Iso1}-\ref{Iso2}. The scenario corresponding to Figures \ref{Iso1a}-\ref{Iso1b} and \ref{Iso2a}-\ref{Iso2b} is depicted in Figure \ref{Case1}; similarly, the scenario corresponding to Figures \ref{Iso1c}-\ref{Iso1d} and \ref{Iso2c}-\ref{Iso2d} is depicted in Figure \ref{Case2} while the scenario corresponding to Figures \ref{Iso1e}-\ref{Iso1f} and \ref{Iso2e}-\ref{Iso2f} is depicted in Figure \ref{Case3}. See the captions below Figures \ref{CasePic}-\ref{Iso2} for the problem scenario descriptions.

First, by comparing the second row to the first row of plots in Figures \ref{Iso1}-\ref{Iso2}, a significant weakening of the electric and magnetic field distributions can be observed. This is caused by the metallic ground's field cancellation effect \cite{balanis1}. The third row of plots in each figure set corresponds to placing the dipole on top of a $d=$5mm thick isoimpedance substrate, with properly chosen material tensors $\boldsymbol{\bar{\epsilon}}_r=\boldsymbol{\bar{\mu}}_r=\mathrm{Diag}[5,5,1/5]$, that mimics the case of a thicker, 25mm free-space buffer separating the source and ground. This leads to a field distribution, for a fixed source-observer depth separation $z-z'>0$, that is exactly identical to that obtained if the source resided in free space 25mm above ground. This result can be also established analytically \cite{teixeira1,pendry1}, and is confirmed numerically upon observing the full agreement between Figures (1) \ref{Iso1a} and \ref{Iso1e}, (2) \ref{Iso1b} and \ref{Iso1f}, (3) \ref{Iso2a} and \ref{Iso2e}, and (4) \ref{Iso2b} and \ref{Iso2f}.

We also make a minor remark concerning the mottled blue annular ``ring'', visible around the central region of intense electric field near the source in Figures \ref{Iso2a}, \ref{Iso2c}, and \ref{Iso2e}. Since the $xy$ plane field distribution cuts are in fact taken one meter above the plane on which the source resides, the $xy$ observation planes intersect the intense main beam of the dipole as well as the deep nulls in the radiation pattern surrounding the main beam. Indeed, observing Figures \ref{Iso1a}, \ref{Iso1c}, and \ref{Iso1e} at the elevation $z-z'\sim1$m, one observes that deep nulls in the dipole's electric field distribution occur at approximately $|x-x'|\sim2$m, which corresponds to the annular region $|\rho-\rho'|\sim$2m in the $xy$ plane electric field plots.
\subsection{Convergence and Accuracy Comparison: CPMWA and CGLQ}
Finally, we perform a study comparing the ability of the CPMWA and CGLQ algorithms to converge to the field contribution due to the evanescent spectra. Since the treatment of the propagation spectra is virtually identical to that in CPMWA, convergence results for the CGLQ algorithm in this latter region are omitted. By demonstrating the CGLQ algorithm's ability to converge to the field contribution from evanescent spectra, we also demonstrate, by extension, the algorithm's ability to converge to the field contribution arising from hybrid spectra. By ``hybrid spectra" we refer to those characteristic plane wave modes that exhibit propagation behavior along the $x$ direction but evanescent behavior along the $y$ direction (or vice-versa); for reference, these two hybrid spectrum regions were denoted Regions IIa and IIb in Figure 3 of \cite{sainath2} and consist of the regions (1) $\left(|Re[k_x]|<\xi_1\right) \cup \left(|\mathrm{Re[}k_y{]}|>\xi_1\right)$ and (2) $\left(|Re[k_y]|<\xi_1\right) \cup \left(|\mathrm{Re[}k_x{]}|>\xi_1\right)$.

In Figures \ref{Conv1}-\ref{Conv2} below, we plot the residual error in the evanescent spectrum field contribution from using the CGLQ and CPMWA algorithms for two representative scenarios. Figure \ref{Conv1} represents a relatively benign scenario, with small transverse source-observer separation $|x-x'|=|y-y'|=1$m and moderate depth separation $z-z'=1$m. In this case, even if $k_x$ and $k_y$ were real-valued along their respective integration paths, the integrand would exhibit low oscillation and fast exponential decay with respect to increasing $|k_x|$ and $|k_y|$. This scenario computes $H_z$ due to a Hertzian (equivalent) magnetic dipole source oriented parallel to the optical axis of a uniaxial medium characterized by the conductivity tensor $\boldsymbol{\bar{\sigma}}=$diag[$\sigma_{xx},\sigma_{yy},\sigma_{zz}$]S/m=diag$[1,1,1/10]$S/m. On the other hand, Figure \ref{Conv2} represents a more challenging scenario if evaluated by standard real-axis integration due to the large $|x-x'|=|\rho-\rho'|=500$m source-observer transverse separation (i.e., $\sim 16.7$ free-space wavelengths) and vanishing $|z-z'|=0$m source-observer depth separation. In this case, we compute $H_y$ radiated by a Hertzian vertical electric dipole in vacuum. Both results are compared against available analytical solutions\footnote{Note that the first scenario admits, as its closed-form solution, the equivalent magnetic dipole fields in an isotropic medium with effective conductivity $\sigma=\sigma_{xx}=\sigma_{yy}=1$S/m \cite{felsen}.}. To illustrate the applicability of the CGLQ algorithm over a wide frequency range, the source radiates at $f=1$kHz in the scenario of Figure \ref{Conv1} and at $f=10$MHz in the scenario of Figure \ref{Conv2}.

For the CPMWA, we vary the Gauss-Legendre quadrature order $P$ used to integrate each of the $B$ extrapolation intervals on a given Fourier integral half-tail, whose successive ``cumulative" integration results\footnote{By ``cumulative" integrals we mean the unprocessed estimates of the non-truncated tail integral obtained by simply integrating over an increasingly longer path \cite{sainath}.} were employed as the input into the CPMWA weighted average computation detailed in \cite{sainath2}. For the CGLQ, we only vary the Gauss-Laguerre quadrature order (also denoted $P$ in the Figures) used to evaluate each Fourier integral half-tail. To facilitate plotting the results, we keep the accuracy of the CGLQ results constant versus increasing $B$ (obviously, there is no integration path splicing in CGLQ).

We observe that in both Figures \ref{Conv1} and \ref{Conv2} the CGLQ algorithm successfully converges to the correct evanescent field spectrum contribution. Not surprisingly, based on results in \cite{sainath2}, the CPMWA also exhibits good convergence characteristics for both scenarios. In Figure \ref{Conv1}, we observe that while CPMWA has a slight better accuracy than CGLQ, the difference is very small. In exchange for this small difference in accuracy, a significant reduction in computational cost is obtained. Observing that the CPMWA method delivers a result with maximum accuracy (relative to the range of $B$ tested and shown in Figure \ref{Conv}) within approximately $B$=6 intervals used for each half-tail path, one realizes that $6 \times 30=180$ integrand evaluations are necessary when using CPMWA; on the other hand, compare this to thirty integrand evaluations using CGLQ. Similarly, a savings factor of about four in computational cost results from comparing CGLQ against CPMWA with $B=6,P=20$ (with a 1-2dB better accuracy exhibited by CPMWA).

In Figure \ref{Conv2} we again notice that both the CPMWA and CGLQ algorithms converge well to the true evanescent field contribution solution, tailing off with a residual error of about -95dB (or approximately nine to ten digits of accuracy) for $B\geq 6$ using either the 20-point or 30-point CGLQ variant and either the 20-point or the 30-point CPMWA variant. Thus, comments concerning computational efficiency gains in this scenario parallel those from the more benign case, with one realizing a factor of four to six in computational cost savings.
\begin{figure}[H]
\centering
\subfloat[\label{Conv1}]{\includegraphics[width=2.9in]{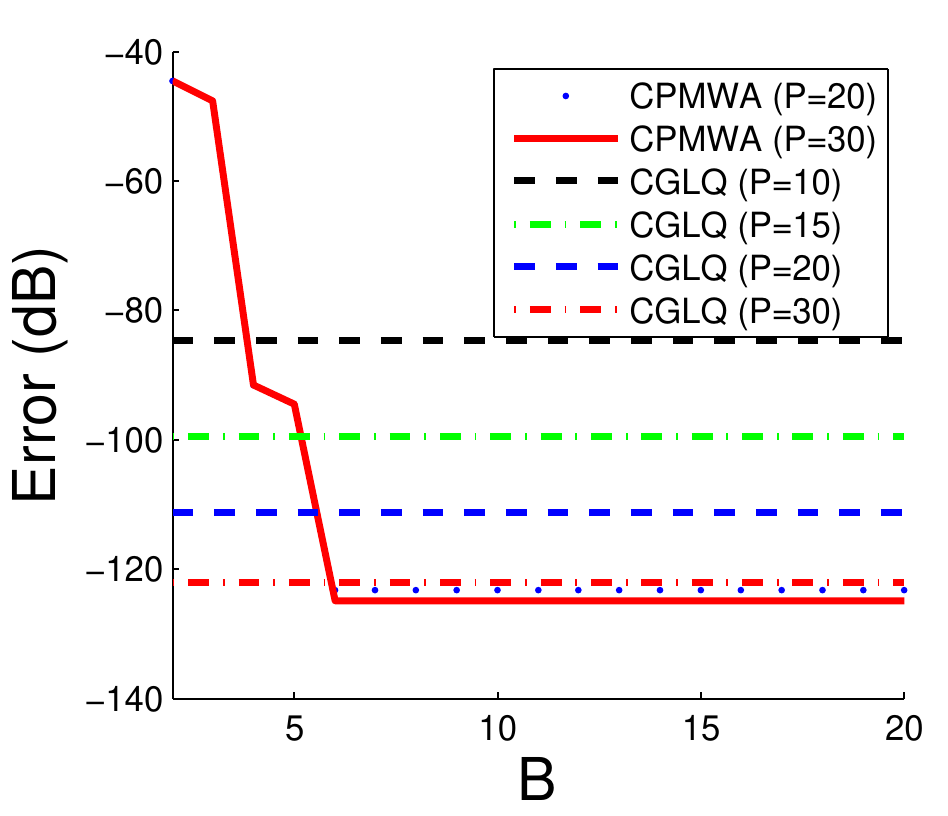}}
\subfloat[\label{Conv2}]{\includegraphics[width=2.9in]{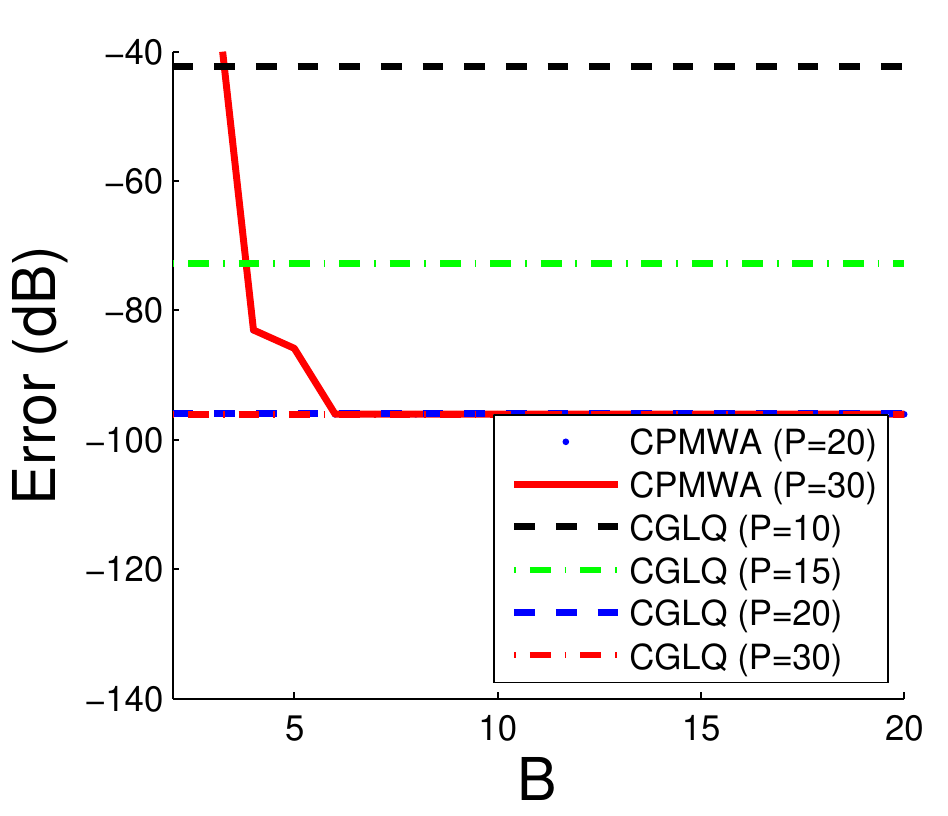}}
\caption{\small Convergence rate and accuracy characteristics for the CGLQ and CPMWA algorithms. To compute the reference evanescent spectrum field contribution values against which the algorithm's results were measured for accuracy, the propagation and hybrid spectrum field contributions were computed with an adaptive integration error tolerance of 1.2d-15, summed together, and subtracted from the closed-form, space domain Hertzian dipole field solution available from \cite{balanis1}.}
\label{Conv}
\end{figure}

\section{Conclusion}
In this work, we have detailed the mathematical formulation behind a novel application of complex-plane Gauss-Laguerre quadrature (CGLQ) to the evaluation of spectral integrals arising in the computation of the tensor Green's function components for planar-stratified media containing layers of arbitrary anisotropy and loss. The proposed CGLQ algorithm touts the ability to robustly guarantee absolute, exponential convergence for the tail integrals for a wide range of frequencies, layer medium properties, source and field type, source orientation, and $\bold{r}-\bold{r}'\neq \bold{0}$ separation geometry. Compared to prior leading algorithms used for this type of problems, the computational burden in computing the hybrid and evanescent spectra has been significantly reduced, computer storage requirements for the numerical quadrature algorithm have also been reduced, algorithm-dependent constraints on the path deformation detour angles have been eliminated, and the numerical instability of weight computations (along with the resultant need for ad-hoc adjustment of the weighted average-type extrapolation schemes) has been eliminated. Furthermore, by replacing the prior CPMWA algorithm's cumbersome interval partition-cum-extrapolation methodology with a highly streamlined process involving one simple Gauss-Laguerre numerical quadrature, the present CGLQ method proves far easier to implement.

To validate the new algorithm's accuracy and convergence properties, two case studies of practical interest involving layered anisotropic media, as well as a convergence study, were performed. The CGLQ algorithm was shown to effect a fast and robust computation of spectral integrals needed for the evaluation of Green's Tensor components in layered anisotropic media. Based on the results shown, we can state that CGLQ stands, at the very least, as a viable competitor to extrapolation-based methods previously touted as the most robust means by which one can robustly compute Fourier-type integrals susceptible to rapid oscillation and small decay rate \cite{mich2,mosig1}. The present contribution has significantly mitigated, in one stroke, both the convergence and computational efficiency bottlenecks associated with the evaluation of the evanescent spectra field contributions that have plagued the direct numerical evaluation of such layered-media Green's Tensor integrals in the past.
\section{Acknowledgments}
This work was supported by a NASA Space Technology Research Fellowship (NSTRF). We acknowledge partial support from the Ohio Supercomputer Center under Grant PAS-0061. We also acknowledge Dr. Burkay Donderici of Halliburton Energy Services and Dr. Anthony Freeman of NASA Jet Propulsion Laboratory for providing helpful impetus for this work.
\section{References}
\bibliographystyle{apsrev4-1}
\bibliography{reflist}
\end{document}